\documentclass[english,journal=jacsat,manuscript=article,etalmode=truncate,maxauthors=0,layout=twocolumn]{achemso}
\setkeys{acs}{etalmode=truncate,maxauthors=0,articletitle = true}

\usepackage{amsmath}             % for equation typesetting
\usepackage{amssymb}             % for equation typesetting
\usepackage{wasysym}             % for geometric shapes
\usepackage{color}               % for colored fonts
\usepackage{setspace}            % for 1.5 and double spacing
\usepackage{graphicx}            % main graphics package
\usepackage{wrapfig}             % allow text wrapping around figures
\usepackage[dvipsnames]{xcolor}  % for inserting colored text
\usepackage{array,booktabs,multirow, multicol}      % for formatting the journal option table at the end of this document

\SectionNumbersOff    % turn off section numbering
\usepackage[font=footnotesize,labelfont=bf,labelsep=period,width=0.95\linewidth]{caption}   % format single-image captions and table titles
\usepackage[font=footnotesize,labelfont=bf,labelsep=period]{subcaption}                     % format subfigure captions
\DeclareCaptionSubType*[arabic]{figure}                                                     % use arabic numerals for subfigure captions (e.g., 1.1, 1.2, etc.)
\DeclareCaptionLabelFormat{subfiglabel}{Figure #2}                                          % append 'Figure' to subfigure captions (e.g., Figure 1.1, Figure 1.2, etc.)
\usepackage[capitalize]{cleveref}
\crefname{figure}{Fig.}{Figs.}
\Crefname{figure}{Figure}{Figures}
\crefname{table}{Tab.}{Tabs.}
\Crefname{table}{Table}{Tables}
\crefname{equation}{Eq.}{Eqs.}
\Crefname{equation}{Equation}{Equations}
\crefname{section}{Sec.}{Secs.}
\Crefname{section}{Section}{Sections}

\title[Running Title]{Chirality-Driven Magnetization Emerges from Relativistic Four-Current Dynamics}

\author{Shiv Upadhyay}
\affiliation[University of Washington]{Department of Chemistry, University of Washington, Seattle, WA 98195, USA}
\alsoaffiliation[co-first]{Authors contributed equally to this work}

\author{Xuechen Zheng}
\affiliation[University of Washington]{Department of Chemistry, University of Washington, Seattle, WA 98195, USA}
\alsoaffiliation[co-first]{Authors contributed equally to this work}

\author{Tian Wang}
\affiliation[University of Washington]{Department of Chemistry, University of Washington, Seattle, WA 98195, USA}

\author{Agam Shayit}
\affiliation[University of Washington]{Department of Physics, University of Washington, Seattle, WA 98195, USA}

\author{Jun Liu}
\affiliation[NCSU]{Department of Mechanical and Aerospace Engineering, North Carolina State University, Raleigh, NC 27695, USA}

\author{Dali Sun}
\affiliation[NCSU]{Department of Physics, North Carolina State University, Raleigh, NC 27695, USA}

\author{Xiaosong Li}
\email{xsli@uw.edu}
\affiliation[University of Washington]{Department of Chemistry, University of Washington, Seattle, WA 98195, USA}
\date{May 2024}

\begin{document}

%-------------------------------------------------------------------------------
% ABSTRACT
%-------------------------------------------------------------------------------
\twocolumn[
\begin{@twocolumnfalse}
\begin{abstract}
Chirality-induced spin selectivity (CISS) is a striking quantum phenomenon in which electron transport through chiral molecules leads to spin polarization---even in the absence of { external} magnetic fields or magnetic components. Although observed in systems such as DNA, helicenes, proteins, and polymers, the fundamental physical origin of CISS remains unresolved. 
Here, we introduce a time-dependent relativistic four-current framework, in which charge and current densities evolve according to the time-dependent variational principle. Real-time relativistic four-current simulations enable direct analysis of helical currents and induced magnetization dynamics. Applied to helicenes---axially chiral molecules lacking stereocenters---our simulations reveal curvature-induced helical electron currents that generate spontaneous magnetic fields aligned along the molecular axis. These fields are handedness-dependent and reach magnitudes of $10^{-1}$~Tesla per single helicene strand.
Our results suggest that CISS may arise from intrinsic, relativistic curvature-induced helical currents and the associated magnetic fields within chiral molecules. This four-current mechanism offers a self-contained explanation for {the driving force underlying spin selectivity, independent of interfacial effects or  unphysically enhanced} spin--orbit coupling. {Furthermore, our results provide a new perspective that offers a unifying framework with the potential to reconcile many existing hypotheses and theoretical models, while also suggesting several testable predictions that can be examined experimentally.}
\end{abstract}
\end{@twocolumnfalse}
]

%\section{Introduction}

Chiral molecules play a vital role in biological processes,\cite{Testa86_60,Freeman94_9203,lo20_7031,Midha01_1205,Cieplak05_1565,Sandars05_49,Mielke14_781,Hore21_043032} and have emerged as promising platforms for engineering spin-dependent technologies.\cite{Rikken11_864,Waldeck12_2178,Sun12_218102,Cuniberti12_081404,Loktev13_165409,Sun14_11658,Waldeck15_263,Naaman16_2560,Waldeck17_1,Waldeck19_250,Paltiel19_133701,Fransson19_7126,Bauer22_1}
Owing to their lack of inversion symmetry, these molecules exhibit a strong coupling to electron spin, manifesting in the phenomenon known as chirality-induced spin selectivity (CISS).\cite{Naaman99_814,Waldeck24_1590}
The CISS effect was first observed in DNA,\cite{Zacharias11_894,Naaman11_4652,Galperin13_13730} where electron transport through chiral structures led to spin-polarized currents even in the absence of an external magnetic field.

Since this pioneering discovery, the CISS effect has been experimentally confirmed in a broad range of systems, including helicenes,\cite{Naaman16_9203,Zacharias18_2025,Schneider22_1,Lingenfelder22_1,Crassous22_7709,Crassous23_1,Mannini23_15189,Burgler24_2308233} proteins,\cite{Fontanesi13_14872,Naaman14_8953,Naaman15_3377,Naaman18_1091,Naaman19_19198,Naaman20_808,Naaman20_20456,Yan20_6607,Naaman21_9875,Naggar23_145101,Waldeck23_6462,Mondal23_024708} and synthetic polymers.\cite{Naaman15_1924,Naaman17_15777,Yan20_6607,Yashima20_15777,Naaman21_7189,Naaman22_2727,Tegenkamp23_17406}
Although numerous theoretical models have been proposed to explain the underlying mechanism,\cite{Mujica15_218102,Medina16_155436,Hasan18_978,Hedegard19_5253,Wees19_024418,Yan21_638} the physical origin of the CISS effect remains a subject of ongoing debate and investigation.

Contemporary theoretical explanations for the CISS effect can be broadly categorized into two classes.
The first viewpoint posits that an effective spin--orbit coupling term must be added to the Hamiltonian describing the electronic structure of a chiral system. However, such a framework struggles to account for the observation of CISS effects in certain hydrocarbon systems, where intrinsic spin-orbit coupling is negligibly small.\cite{Kronik22_2106629,Thijssen23_6900}

The second viewpoint attributes spin selectivity primarily to interface effects between electrodes or substrates and the chiral molecules. In this class of theories, spin--orbit interactions in the substrate convert the orbital angular momentum, filtered by the chiral molecule into spin angular momentum, resulting in spin-polarized transmission.\cite{Boulougouris13_114111,Yan21_638,Dubi22_10878,Herrmann20_7357} While this mechanism offers a plausible explanation for surface-bound systems, it fails to explain recent gas-phase experiments,\cite{Wasielewski23_197} where clear CISS effects were observed in the complete absence of electrodes or supporting substrates.

These inconsistencies highlight the need for a more general and intrinsic explanation of the CISS phenomenon---one that captures the fundamental role of chirality in shaping spin dynamics, independent of external interfaces or large spin--orbit couplings. In this context, a fully \emph{ab initio} relativistic electron dynamics based on the four-current formalism may offer a promising path forward.

Accurately describing electron dynamics in atoms, molecules, and materials remains a central challenge in modern theoretical chemistry and physics.
Real-time time-dependent density functional theory (RT-TDDFT)\cite{Li05_084106,Rubio06_2465,Li07_134307, Li11_184102,Gray11_196806,Govind11_1344,Govind12_3284,Li14_214111,Isborn15_4791,Li16_234102,Li16_104107,Li18_1998,Li19_234103} and time-dependent wave function approaches, such as time-dependent coupled-cluster (TD-CC)\cite{Schlegel11_4678,HeadGordon12_909,Jorgensen15_114109,Jorgensen16_024102,DePrince16_5834,DePrince17_2951,Li19_6617,Li21_5438,Li23_044113} and time-dependent configuration interaction (TD-CI)\cite{Levine15_4708,Levine15_024102,Li18_295,Levine18_4129,Wilson18_014107,Li19_1633}, have enabled practical simulations of electronic excitations and real-time electron dynamics.\cite{Li18_e1341, Li20_9951}
However, conventional formulations of these methods often neglect essential relativistic and gauge-covariant aspects critical to the accurate modeling of current-carrying systems. A fully consistent and complete description of electron dynamics requires adopting the four-current formalism, which includes both the charge density and the three-component current density vector.

The current density is the most important quantum mechanical property that underlies the magnetic response and dynamics of the electrons, including NMR spectroscopy, magnetic susceptibilities, spin transport, and magnetically induced current pathways in molecules and materials.\cite{Nitzan11_2118,Nitzan12_155440,Berger21_12362,Garner25_4073,Hod20_8652} 
Within the Schr\"odinger framework, current can be introduced as a velocity density.\cite{Gauss04_3952,Sundholm11_20500} However, the nonrelativistic velocity density fails to  capture spin-dependent currents or those arising from spin-couplings (\emph{e.g.}, spin--orbit, spin--spin), such as those described by the Dirac--Coulomb--Breit operator.\cite{Li21_3388,Li22_064112,Li23_171101}
The full treatment of the four-current within the four-component relativistic framework was developed by Saue and co-workers for studies of magnetically induced current density,\cite{Saue09_187} and was later extended to the London orbital framework.\cite{Saue11_20682}

\begin{figure}[H]
    \centering
    \includegraphics[width=\linewidth]{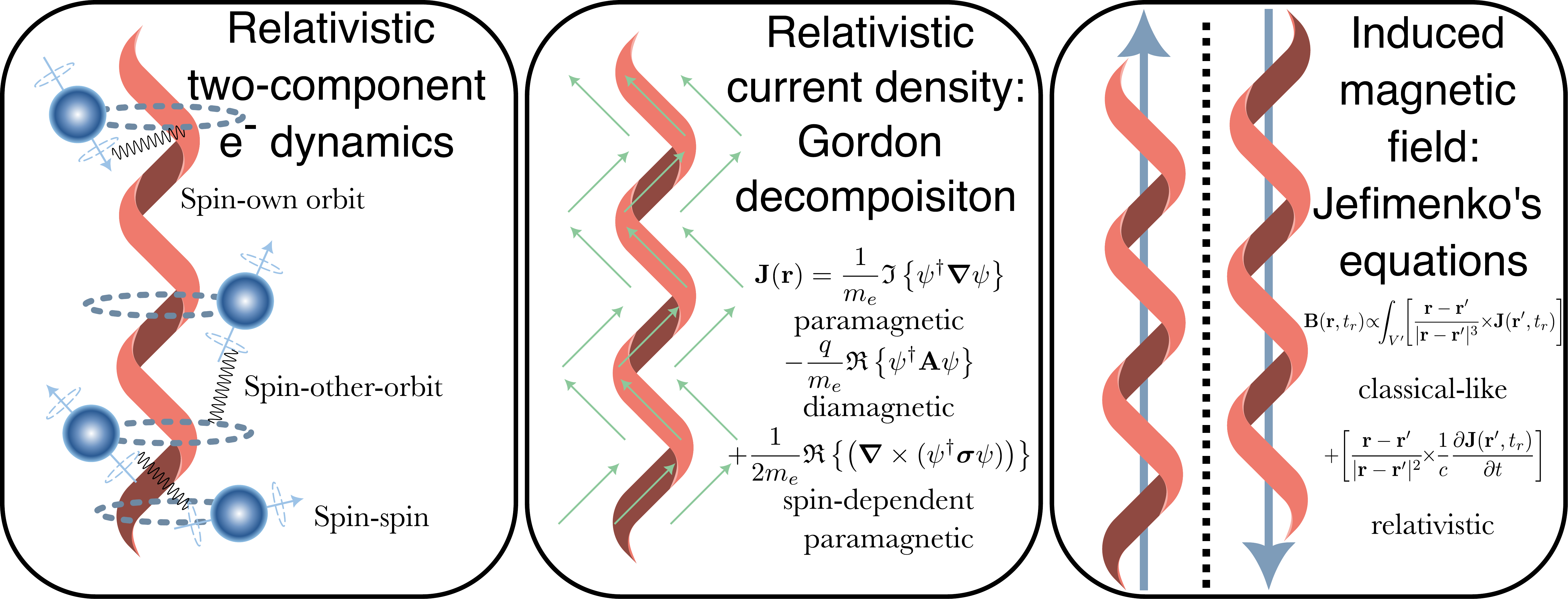}
    \caption{Simulation workflow. The relativistic two-component TDDFT equations are propagated in real time, generating the four-current density on-the-fly via the Gordon decomposition. Integration of the resulting four-current reveals a spontaneous magnetization arising from chiral, curvature-induced helical currents.}
    \label{fig:workflow}
\end{figure}

In this work, we present a time-dependent relativistic four-current framework in which both charge and current densities evolve according to the time-dependent variational principle.
\Cref{fig:workflow} illustrates the workflow of the relativistic four-current dynamics approach. The simulation begins with relativistic two-component RT-TDDFT electron dynamics. The resulting two-component density is then transformed into the four-current formalism using the Gordon decomposition. Spatial integration of the four-current yields an effective magnetic field.

\begin{figure}[H]
    \centering
    \includegraphics[width=\linewidth]{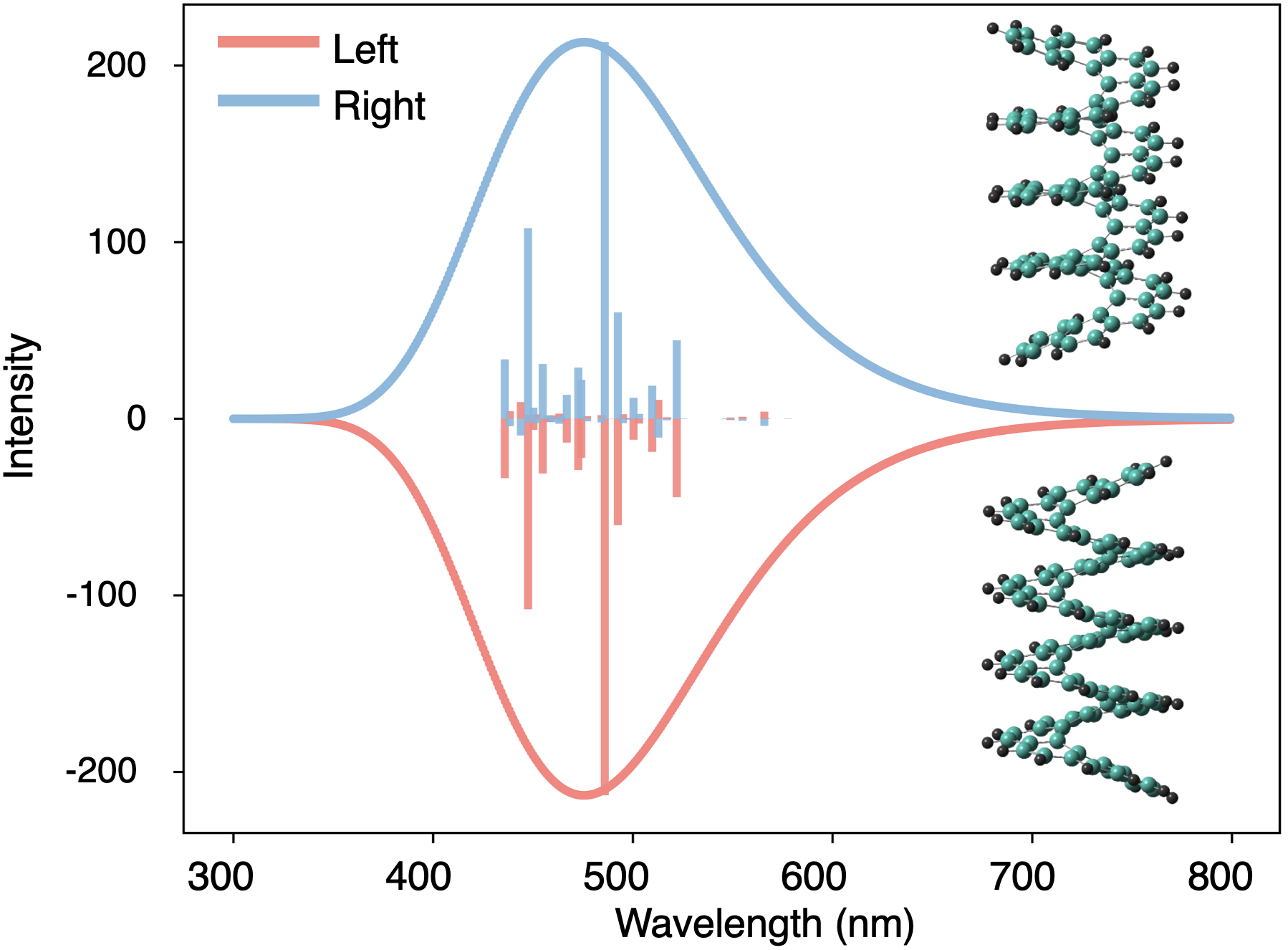}
    \caption{Computed ECD spectra of (L)-[24]helicene and (R)-[24]helicene.
    }
    \label{fig:24helicenecd}
\end{figure}

\begin{figure}[H]
    \centering
    \includegraphics[width=\linewidth]{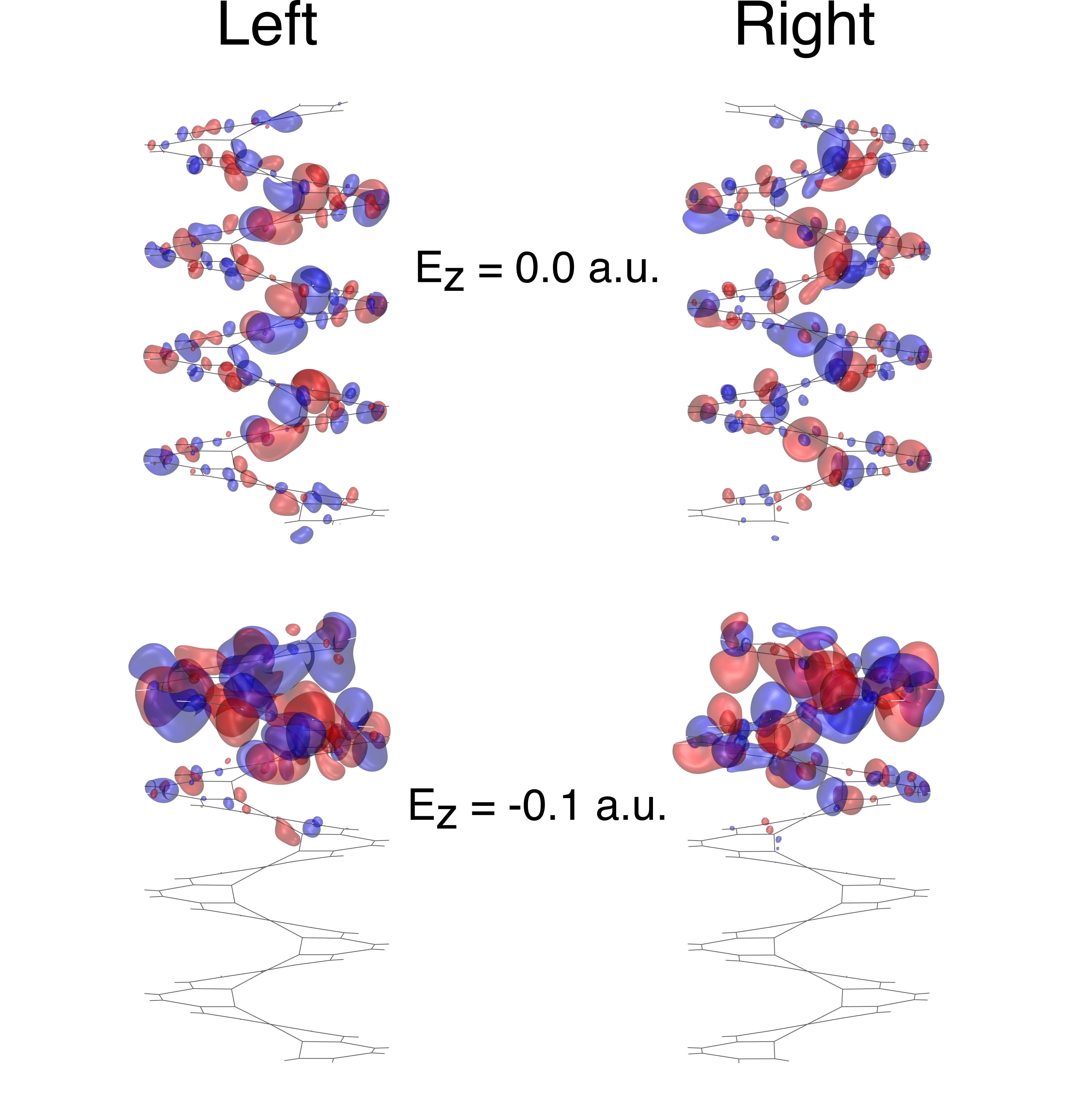}
    \caption{Highest singly occupied molecular orbital of anionic [24]helicene: ({\bf top}) field-free ground state and ({\bf bottom}) under a 0.01 au electric field aligned along the $-z$-direction.
    }
    \label{fig:24helicenescf}
\end{figure}

We apply this framework to investigate CISS effects in a helicene system using real-time relativistic exact-two-component RT-X2C-TDDFT quantum dynamics. Although helicenes do not possess chiral carbon stereocenters, they exhibit axial chirality, which plays a crucial role in determining their electronic dynamics. As illustrated in \cref{fig:24helicenecd}, the computed electronic circular dichroism (ECD) spectra confirm the chiral nature of L- and R-[24]helicene.

\begin{figure*}
    \centering
    \captionsetup{width=.9\linewidth}
    \includegraphics[width=0.9\linewidth]{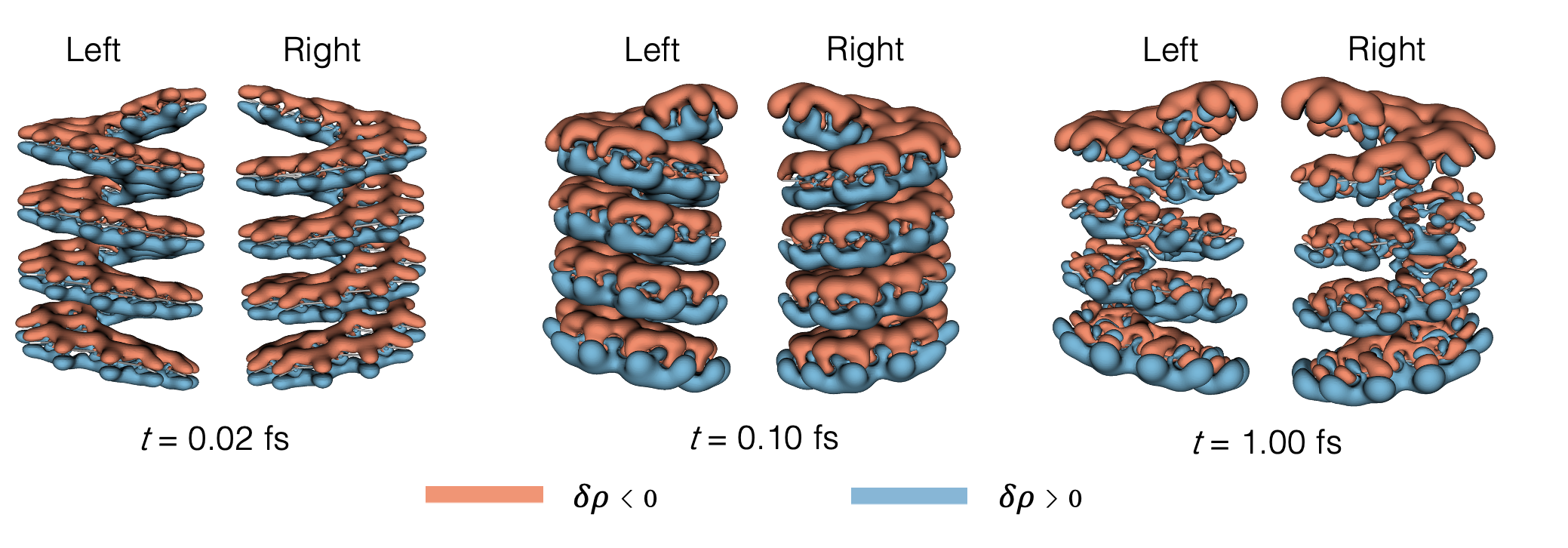}
    \caption{Time-resolved charge density, shown as the difference $\delta\rho_\text{s} = \rho_\text{s}(t) - \rho_\text{s}(0)$, at selected time points for (L)-[24]helicene and (R)-[24]helicene.
    }
    \label{fig:densitydiff}
\end{figure*}

{Specifically, we aim to simulate the experimental condition in which a helical system is subjected to an electric field bias, resulting in electron transport through the helicene.\cite{Naaman16_9203,Schneider22_1,Lingenfelder22_1,Crassous22_7709,Crassous23_1,Mannini23_15189,Burgler24_2308233}}
For the electron transport simulations, an extra electron is added to the helicene molecule, forming an anionic molecular system. {The introduction of a spin-1/2 system with an extra electron is specifically designed to probe the spin-current contribution, which can be rigorously described within the relativistic four-current formalism.} 

The spatial distribution of this additional electron in the ground state is shown in \cref{fig:24helicenescf}. In the absence of an external electric field, the electron is delocalized across the entire molecule, with the highest density concentrated near its center. The relativistic RT-X2C-TDDFT electron dynamics are initiated from an electric-field-polarized state (see Theory and Computational Details for more information), with the initial molecular orbital depicted in \cref{fig:24helicenescf}. The quantum dynamics are driven by an external field $E_z = -0.01$ au in the $-z$ direction. {The total simulation time is 1 fs, and a complex absorbing boundary condition was applied.\cite{Kummel16_035115} During the simulation, the electrons did not reach the boundaries; therefore, no reflection occurred, and we observed no significant difference between the scenarios with and without the complex absorbing boundary condition. This setup represents a non-equilibrium transport condition.\cite{DiVentra05_2569,VanVoorhis06_155112,Kummel16_035115} }

Relativistic four-current dynamics simultaneously capture the time-dependent behavior of both charge $\rho_\text{s}$ and current  ($J_x$, $J_y$, $J_z$) densities. We begin by examining the evolution of the charge density. \Cref{fig:densitydiff} presents the time-resolved charge or scalar density, shown as the difference density $\delta\rho_\text{s} = \rho_\text{s}(t) - \rho_\text{s}(0)$, at selected time points. While the charge density evolution reflects a typical electron transfer process, the dynamics indicate that the electron density remains confined to the helicene backbone. This suggests that the helical curvature of the structure plays a key role in shaping the characteristics of charge transport.

\begin{figure}[H]
    \centering
    \includegraphics[width=\linewidth]{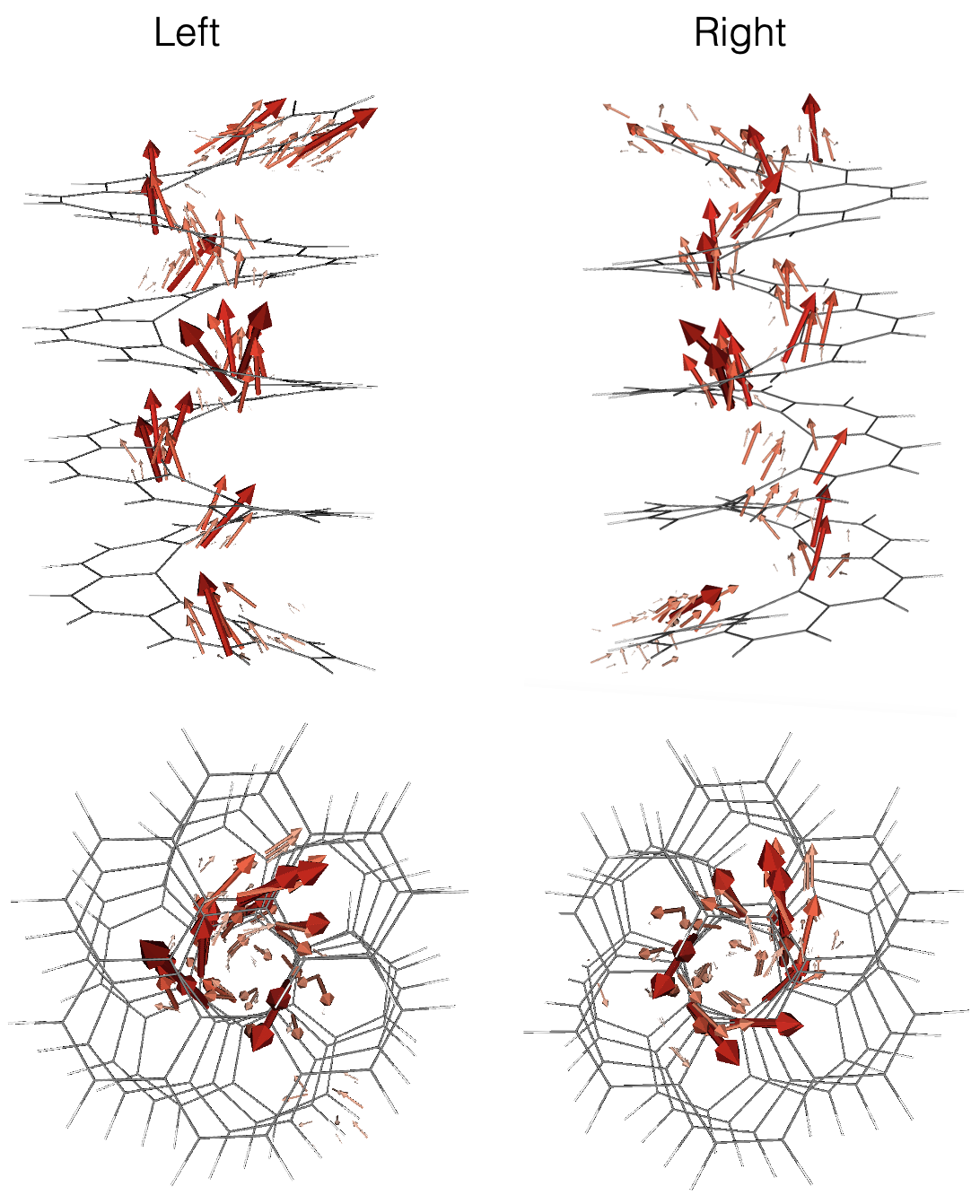}
    \caption{Side and top views of the helical current in (L)-[24]helicene and (R)-[24]helicene at 0.10 fs.}
    \label{fig:helicene24J}
\end{figure}

The helical curvature of the molecule can induce a rotational component in the electron current. This effect is demonstrated in \cref{fig:helicene24J}, where the vector form of the electron current, represented by (${J}_x(\mathbf{r},t)$, ${J}_y(\mathbf{r},t)$,  ${J}_z(\mathbf{r},t)$), is shown as a function of space at $t=0.1$ fs when the helical-current is at its maximum at the center of the helicene. Rather than exhibiting a simple axial velocity current along the $z$-direction, the current vector displays a pronounced rotational behavior that tracks the helical curvature of the structure, as seen in \cref{fig:helicene24J}. This rotation highlights the direct influence of molecular chirality on current dynamics.

To examine the current vector in more detail, \cref{fig:helicene24J} displays the current direction in the plane perpendicular to the chiral axis (the 
$z$-direction). As shown, (L)-[24]helicene produces a counterclockwise current, whereas (R)-[24]helicene generates a clockwise current when viewed from above, looking down along the $-z$-axis. We would like to emphasize that in both simulations, the direction of charge transfer is identical, specifically, from 
$+z$ to $-z$. This suggests that the observed difference in current circulation arises solely from the molecular handedness.

\begin{figure}[H]
    \centering
    \includegraphics[width=\linewidth]{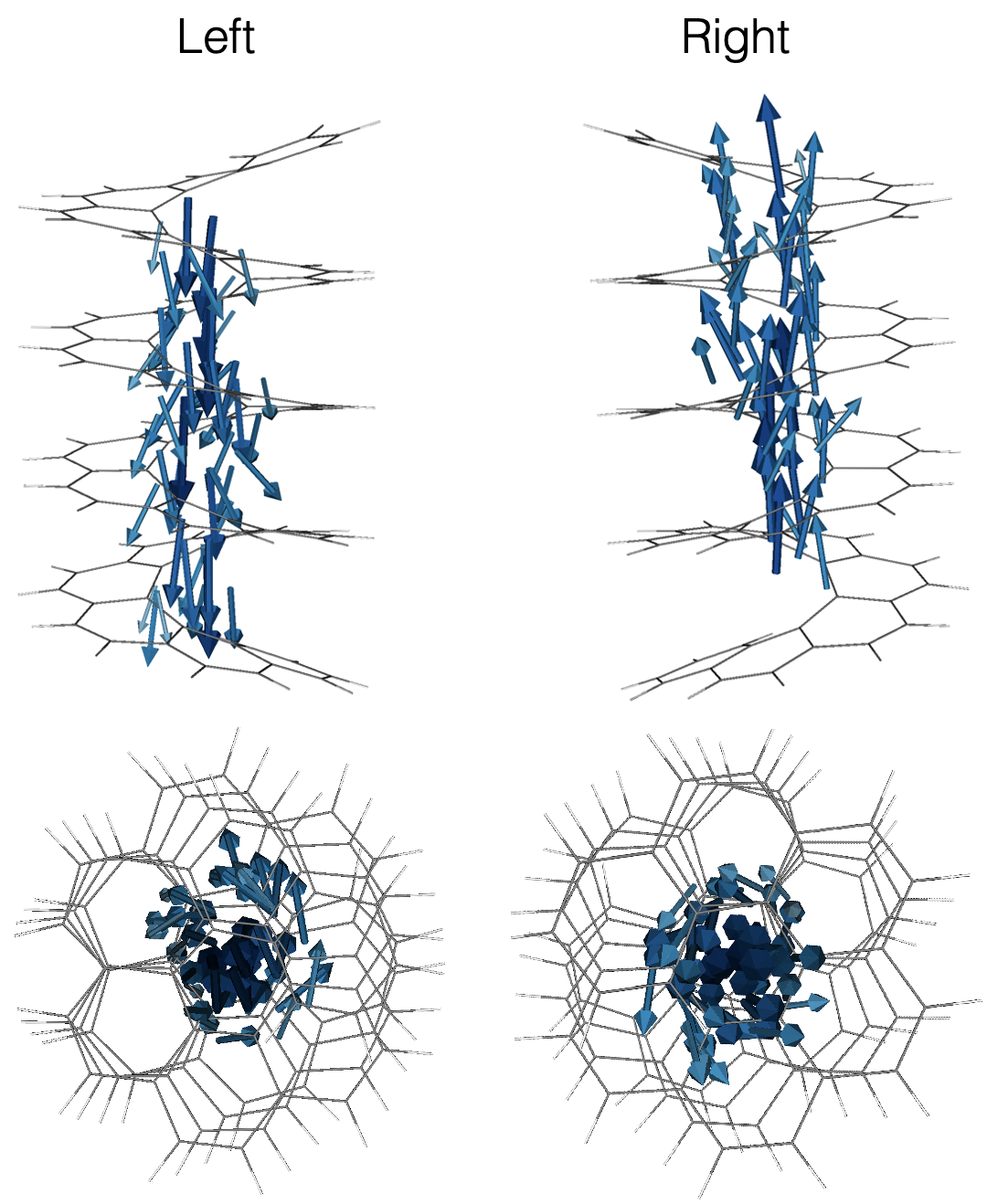}
    \caption{Side and top views of the helical-current-induced magnetic field in (L)-[24]helicene and (R)-[24]helicene at 0.10 fs.}
    \label{fig:helicene24B}
\end{figure}
According to classical electrodynamics, helical currents within a molecule generate a magnetic field whose direction can be determined using the right-hand rule. In the case of (L)-[24]helicene and (R)-[24]helicene, the handedness of the molecule causes the induced magnetic fields to point in opposite directions.
Within the \emph{ab initio} framework, this emergent magnetic field can be computed by integrating the current density using Jefimenko’s equations (see the Theory and Computational Details section for details). \Cref{fig:helicene24B} illustrates the emergence of a magnetic field induced by the progression of helical curvature-induced currents. Notably, (L)-[24]helicene and (R)-[24]helicene produce magnetic fields of similar magnitude but opposite direction, reflecting their mirror-image structures.

\begin{figure}[H]
    \centering
    \includegraphics[width=\linewidth]{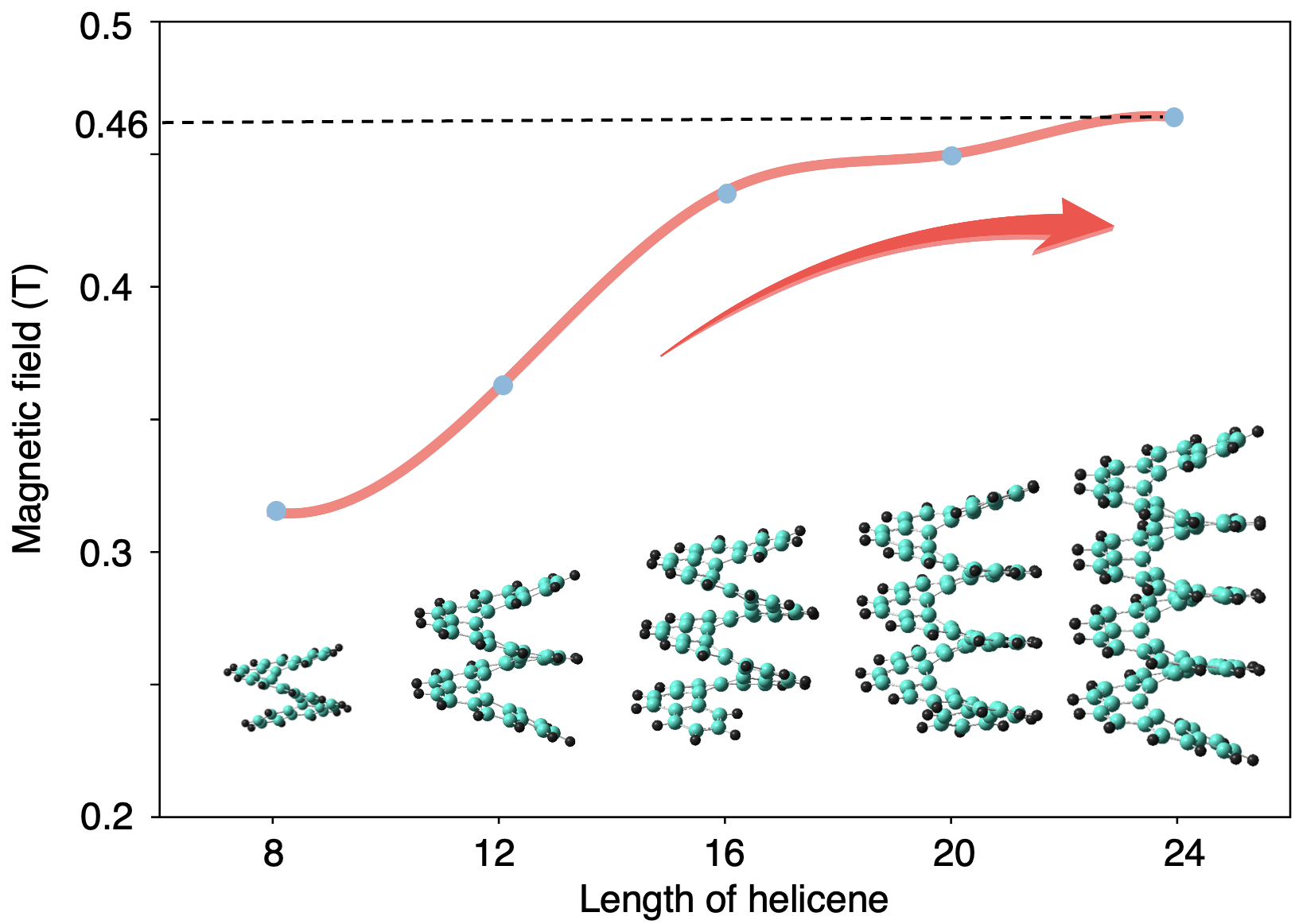}
    \caption{The maximum generated helical-current-induced magnetic field for a variety of helicene lengths.}
    \label{fig:magfieldwaterfall}
\end{figure}

To investigate the length dependence of the helical-current-induced magnetic field, a series of helicenes with varying heights were simulated using four-current dynamics. The maximum magnetic field generated for each helicene is shown in \cref{fig:magfieldwaterfall}. The field strength is $\sim10^{-1}$ Tesla. {As shown in the figure, the magnetic field increases with helicene height and reaches a plateau beyond [16]helicene.} This trend suggests the presence of a critical helical length beyond which the helical-current-induced magnetic field saturates. This behavior can be attributed to the finite delocalization length of $\pi$-conjugated electrons, which governs low-energy electron transport in such systems.

\begin{figure}[H]
    \centering
    \includegraphics[width=\linewidth]{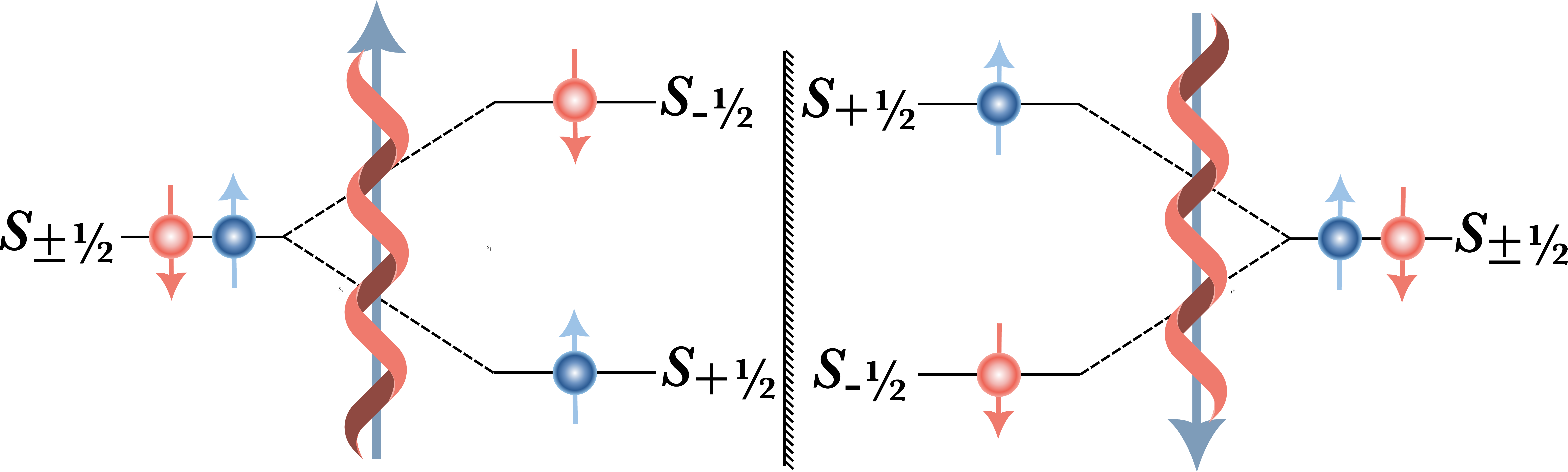}
    \caption{A schematic representation of the helical-current-induced magnetic field giving rise to broken spin (Kramers') symmetry.}
    \label{fig:split}
\end{figure}

The emergence of the helical-current-induced magnetic field is a likely physical driving force underlying the CISS effect. A magnetic field can break spin (Kramers') symmetry, lifting the degeneracy between spin-up and spin-down states (see \cref{fig:split} for an illustration). The direction of the field determines which spin manifold is energetically favored, thereby enabling a spin-filtering effect. 

{From the Boltzmann population analysis of the Kramers' pair (spin-down and spin-up) at 5 K under a 0.46 T magnetic field generated by the helical current in [24]helicene, the spin-down/spin-up ratio is approximately 0.9, indicating that the magnetic field favors one spin manifold by about 10\%. However, this small magnitude suggests that the helical-current-induced magnetic field, although able to break time-reversal symmetry, is by itself insufficient to account for the full extent of the CISS effect. Additional mechanisms must therefore exist to amplify the magnetic field induced by the helical current.}

\begin{figure}[H]
    \centering
    \includegraphics[width=\linewidth]{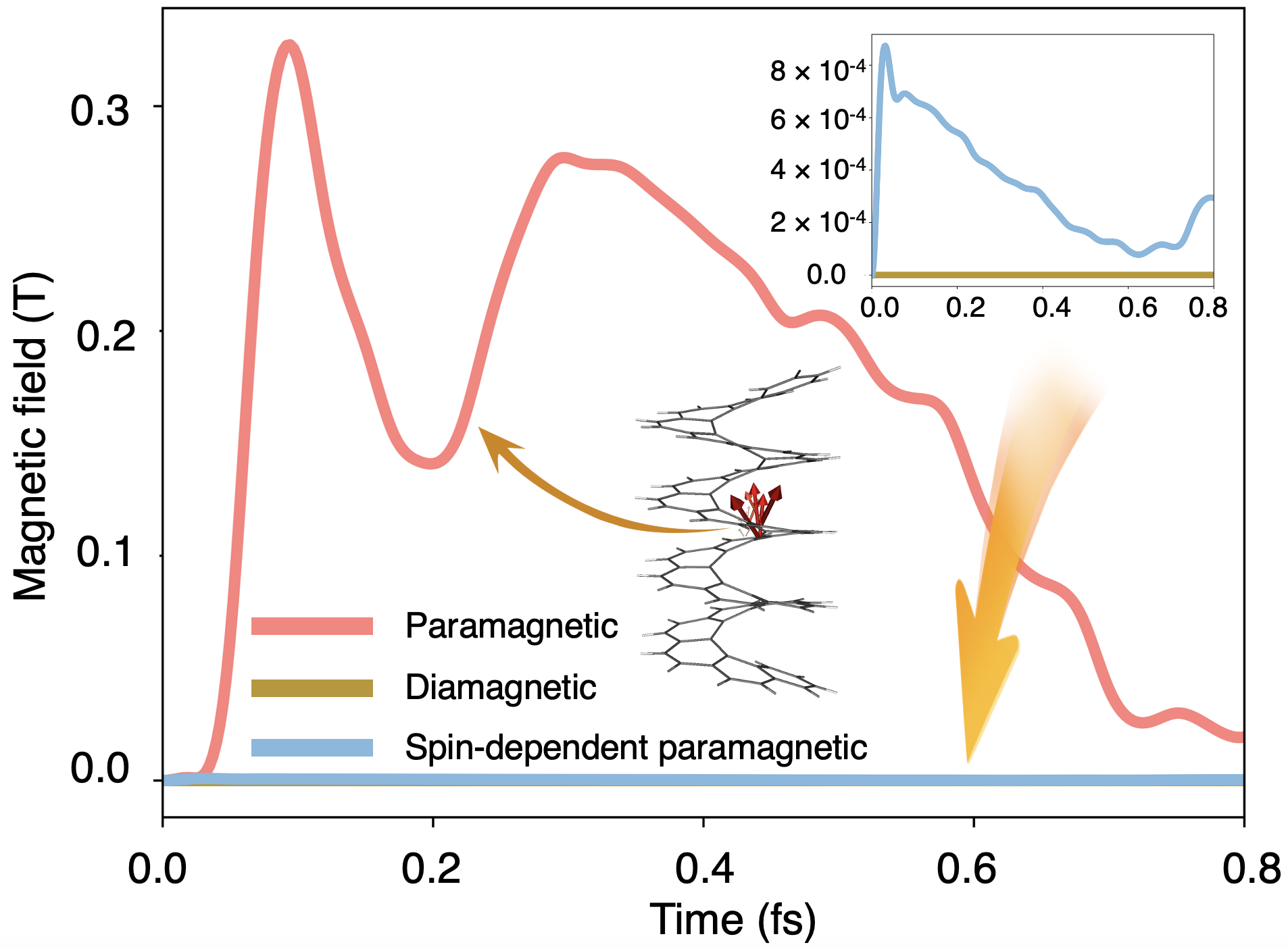}
    \caption{Contributions of paramagnetic, diamagnetic, and spin-dependent paramagnetic terms to the magnetic field induced by helical currents. The inset highlights the spin-dependent paramagnetic contribution. The helicene structure is shown with current arrows indicating the point where the time-dependent magnetic field is evaluated.}
    \label{fig:helicene24J2}
\end{figure}

The four-current expression comprises three terms, each offering key insights into the nature of the helical current. As shown in \cref{fig:helicene24J2}, the velocity current (first term in \cref{eq:gordonx2c}) dominates, being approximately two to three orders of magnitude larger than the spin-dependent paramagnetic current (third term). The diamagnetic term (second term) is absent due to the lack of an external vector potential. Both the velocity and spin-dependent currents originate from the molecular helical curvature; the latter additionally depends on the intrinsic angular momentum of the system.

{From the perspective of a foundational theoretical framework, several important conclusions can be drawn. The velocity current (first term) can be fully accounted for within non-relativistic quantum dynamics, indicating that the observed spontaneous generation of the magnetic field has a non-relativistic origin. Although the spin-dependent paramagnetic current (third term) requires a fully variational relativistic treatment, its contribution in the helicene system is negligible, indicating that a self-consistent treatment of spin and spin–orbit interactions is not necessary in this case. However, in systems with strong spin and spin–orbit couplings, the spin-dependent paramagnetic current can become significant, demanding the use of relativistic four-current dynamics.  A fully self-consistent treatment of the diamagnetic term (second term) requires a quantum field description, which could be approximated using a semiclassical approach in future work.}

\section*{Conclusion}

The physical origin of chirality-induced spin selectivity (CISS) effects in helicenes was investigated using relativistic four-current dynamics within the RT-X2C-TDDFT framework. Simulations revealed a curvature-induced helical current arising from the electron dynamics in the helicene structure. Integration of this helical current over the transport region generates a spontaneous magnetic field aligned with the helical axis. Although the direction of electron transport remains the same in both L- and R-helicenes, the direction of the resulting helical-current-induced magnetic field depends on the molecular handedness, providing compelling evidence for a chirality-induced magnetic response.

The computed helical-current-induced magnetic field for a single helicene lies in the range of $10^{-1}$ Tesla. Its magnitude increases with the length of the helicene, reaching a maximum at a critical threshold length, around [16]helicene, beyond which further increases in length do not significantly enhance the field.

{This work examines the origin of spontaneous magnetization arising from helicity-induced electron currents. As shown in \cref{eq:jefimenko_b}, if the four-current does not have a preferential axis or is not associated with a principal axis of electron transport, the isotropic averaging of the induced field vanishes, resulting in no net magnetization. In contrast, when electron transport occurs along a well-defined direction in a helical structure, the resulting preferential alignment of the four-current breaks isotropic cancellation and gives rise to a non-vanishing spontaneous magnetization. This highlights the fundamental connection between helicity, directional electron motion, and the breaking of isotropic symmetry, which together underpin the emergence of magnetization in chiral systems.} The helical-current-induced magnetic field is likely the origin of the CISS effect by breaking spin symmetry and favoring one spin orientation, enabling spin filtering.

{While the system studied here is specifically tied to helicity-induced electron currents, the four-current quantum dynamics framework is completely general. It enables the analysis of a broad range of driving forces in diverse physical and chemical contexts, including systems with intrinsic chiral centers, strong spin–orbit interactions, and other symmetry-breaking mechanisms.}

\section*{Perspective and Hypotheses}

{
The physical insights gained from this work provide a new perspective on the CISS effect. While the helical-current-induced magnetization accounts for spin Zeeman splitting, it must be amplified to explain the full extent of spin selectivity. This consideration introduces the concepts of a `driving force' and `amplification' in CISS, offering a unifying framework with the potential to reconcile many of the existing hypotheses and theoretical models. The `driving force' is the helical current, which may arise from electron transport through a helical structure and/or from chiral phonons, the latter also correlating with temperature effects. Potential `amplification' mechanisms include strong spin–orbit coupling, which enhances spin-dependent paramagnetic contributions; molecular aggregation, which increases the overall current; and quantum-field self-consistent interactions, which strengthen the diamagnetic response.
}

Together, these analyses and simulation results form the foundation for informed hypotheses regarding the physical origin of the CISS effect:

\begin{itemize}
\item The low-energy conduction band electrons in organic helical systems possess a finite delocalization length. Once this threshold is exceeded, additional $\pi$-conjugation no longer contributes effectively to the helical-current-induced magnetic field.
\item The helical pitch and radius determine the curvature-induced helical current. Therefore, molecular engineering of these structural parameters offers a viable route to modulate the magnitude of the helical current.
\item Although a single helical strand can generate a chirality-dependent magnetic field in the range of $10^{-2}$ to $10^{-1}$ Tesla, this field is likely too weak to fully account for complete spin filtering. Potential amplification mechanisms include strong spin--orbit coupling within the molecule or at the electrode interface, as well as the formation of ordered aggregates to enhance current density.
\item The helical-current-induced magnetic field can, in turn, enter the expression for the helical current itself  as a vector potential in the second term of  \cref{eq:gordonx2c}. Consequently, a self-consistent treatment of the field–current interaction may further amplify the observed magnetic field. {Ideally, these effects would be rigorously treated within a quantum field framework, providing motivation for the development of four-current quantum dynamics coupled with quantum-field-enabled time-dependent electronic structure theory.}
\item In insulators, mechanistic driving forces, such as chiral phonons with non-zero momentum, can transport four-current densities through a helical structure, thereby generating helical currents and spontaneous magnetic field.
\end{itemize}

\section*{Methods}
\subsection*{Theory and Computational Details}

In this paper, the following notation is adopted, unless otherwise specified:
\begin{itemize}
    \item $\{\mu,\nu,\lambda,\sigma\}$ : atomic orbitals (AOs)
    \item $\varphi_i(\mathbf{r},t), i \in {p,q,r,s}$ : molecular orbitals (MOs)
\end{itemize}

\subsection*{Relativistic RT-X2C-TDDFT}
In this work, we use the exact-two-component (X2C) transformation to yield an electron-only relativistic Hamiltonian.\cite{Dyall97_9618,Dyall98_4201,Enevoldsen99_10000,Dyall01_9136,Liu05_241102,Peng06_044102,Cheng07_104106,Saue07_064102,Peng09_031104,Liu10_1679,Liu12_154114,Reiher13_184105,Li16_3711,Li16_104107,Repisky16_5823,Li17_2591,Cheng21_1536,Li22_2947,Li22_2983,Li22_5011,Saue22_114106}
Specifically, we use the one-electron X2C transformation, which allows for one-step approach to construct a unitary transformation matrix to ``fold'' the small component coefficients into the large component. The one-electron X2C approach makes use of the effective one-electron spin--orbit Hamiltonian and avoids the four-component self-consistent-field procedure. In this work, we use the new Dirac--Coulomb--Breit-parameterized effective one-electron spin--orbit Hamiltonian in the X2C approach.\cite{Li23_5785} {The relativistic Hamiltonian does not include explicit quantum field operators; therefore, the four-current formalism employed in this work should be regarded as an effective quantum electrodynamics treatment. It is important to note that the perturbative Pauli Hamiltonian is not bounded from below and, therefore, is unsuitable for describing relativistic four-current dynamics.}

The iterative solution of the two-component self-consistent-field equations yields a set of complex-valued molecular spinors, 
\begin{equation}
    \psi_{p}(\mathbf{x},t) = 
    \begin{pmatrix}  
     \varphi_{p}^{\alpha} (\mathbf{r},t) \\
      \varphi_{p}^{\beta} (\mathbf{r},t) \\
    \end{pmatrix} 
    = 
    \begin{pmatrix}  
      \sum_{\mu}C_{\mu p}^{\alpha}(t)\chi_{\mu}(\mathbf{r}) \\
      \sum_{\mu}C_{\mu p}^{\beta}(t)\chi_{\mu}(\mathbf{r})
    \end{pmatrix} ,\label{eq:spinor}
\end{equation}
where $\mathbf{x}$ is a collective coordinate of both spatial $\{\mathbf{r}\}$ and spin coordinates $\{\alpha,\beta\}$.
The spatial functions $\varphi_{p}^{\{\alpha,\beta\}} (\mathbf{r},t)$ are a linear combination of real-valued atomic orbitals. $C_{\mu p}(t)$ are the time-dependent coefficients.

%The time-dependent Schr{\"o}dinger equation :
%\begin{equation} 
%    i \hbar \frac{\partial}{\partial t} \Psi(t) = \hat{H}(t) \Psi(t)
%\end{equation}
%where $\hat{H}(t)$ is the time-dependent Hamiltonian. 
Time-dependent density functional theory (TDDFT) can be formulated and evolved using the Liouville von Neumann equation in matrix form,\cite{Li18_e1341,Li20_9951}
\begin{equation}
    i\frac{\partial \mathbf{P}(t)}{\partial t}=[\mathbf{F}(t), \mathbf{P}(t)]
\end{equation}
where $\mathbf{F}$ and $\mathbf{P}$ are the Fock/Kohn--Sham and the density matrices in orthonormal basis. 

For two-component real-time methods, the density matrix can be represented in spin-blocked form or as a sum of the scalar density $\mathbf{P}_s$ and the vector magnetization density $\mathbf{P}_x,\mathbf{P}_y,\mathbf{P}_z$:
\begin{align}
\mathbf{P} &= 
\begin{pmatrix}
\mathbf{P}^{\alpha\alpha} & \mathbf{P}^{\alpha\beta} \\
\mathbf{P}^{\beta\alpha}  & \mathbf{P}^{\beta\beta}
\end{pmatrix} \label{eq:spin-block} \\
&=  \mathbf{P}_s\otimes \mathbf{I}_2+ \sum_{k\in\{x,y,z\}}\mathbf{P}_k\otimes\boldsymbol{\sigma}_k \label{eq:magdensity}\\
\mathbf{P}_s &= \frac{1}{2}(\mathbf{P}^{\alpha\alpha}+\mathbf{P}^{\beta\beta})\notag\\
\mathbf{P}_x &= \frac{1}{2}(\mathbf{P}^{\alpha\beta}+\mathbf{P}^{\beta\alpha})\notag\\
\mathbf{P}_y &= \frac{i}{2}(\mathbf{P}^{\alpha\beta}-\mathbf{P}^{\beta\alpha})\notag\\
\mathbf{P}_z &= \frac{1}{2}(\mathbf{P}^{\alpha\alpha}-\mathbf{P}^{\beta\beta})\notag
\end{align}
where $\boldsymbol{\sigma}=(\sigma_x,\sigma_y,\sigma_z)$ is the vector of Pauli matrices. 
%The elements of the density matrix $\mathbf{P}$ are given in the atomic basis by 
%\begin{equation}
%    P_{\mu\nu}(t) = \sum_{p}f_{p}C_{\mu p}(t)C_{\nu p}^{\ast}(t)
%\end{equation}
%where $f_{p}$ is the occupation of orbital $p$.

Commonly used density functional approximations that only depend on the spin-collinear densities ($\mathbf{P}^{\alpha\alpha}$ and $\mathbf{P}^{\beta\beta}$) must be modified to be compatible with two-component Hamiltonians.\cite{Gyorffy01_206403,vanWullen02_779,Liu03_597,Engel04_012505,Ziegler04_12191,Liu04_6658,Liu05_241102,Liu05_054102,Baerends05_204103,Saue05_187,Peng06_044102,Ziegler07_5270,Frisch07_125119,Saue09_2091,Autschbach09_194102,Frisch12_2193,Scuseria13_035117,Weigend13_5341,Liu13_3741,Gross13_156401,Vignale13_245102,Li17_2591} 
The non-collinear DFT framework used in this work redefines the functionals (in the similar spirit of spin-polarized DFT) to depend on a set of auxiliary generalized variables  which take the full magnetization vector density into account,\cite{Frisch07_125119,Frisch12_2193,Scuseria13_035117,Li17_2591}
As this approach has been developed and calibrated, we refer readers to Ref. \citenum{Li17_2591} for mathematical derivations and implementations and Ref. \citenum{Li18_e25398} for a recent review of non-collinear DFT methods.

%{ When applied to sufficient large systems, RT-TDDFT can be used to simulate steady-state electronic transport.\cite{DiVentra05_2569,VanVoorhis06_155112,Kummel16_035115} Compared to contemporary methods to simulate steady-state transport such as nonequilibrium Green’s function (NEGF) methods, RT-TDDFT is extremely attractive since it provides a more granular, all-electron description of electron transport. Additionally RT-TDDFT can be used to simulate more realistic, i.e. more complicated, chemical systems where the boundary between electrode and scattering regions becomes blurred. Furthermore, NEGF often ignore spin-dependent interactions, which are vital for phenomena such as the CISS effect, since the inclusion of spin--orbit coupling beyond a scalar description in NEGF methods is nontrivial.\cite{Wittemeier25_arxiv} When applying RT-TDDFT to simulate electronic transport, a sufficiently large system must be simulated for several femtoseconds to approach the open quantum system steady-state transport limit.\cite{DiVentra05_2569,VanVoorhis06_155112} These constraints combined with the added complexity of relativistic electronic structure has meant that until recently relativistic electron transport simulations were generally considered out of reach. }

%%%%%%%%%%%%%%%%%%%%%%%%%%%%%%%%%%%%%%%%%%%%%%%%%%%%%%%%%%%%%%%%%%%%%%%%%%%%%%%%%%%%%%%%%%%%%%%%%%%%%%%%%%%%%%%%%%%
\subsection*{Two-Component Four-Current}

While the two-component density is convenient for implementing the time-dependent variational principle, its corresponding four-current representation offers a more physically meaningful description in capturing the interplay between charge, current, and spin dynamics, particularly in relativistic and magnetically active systems.

The Gordon decomposition of the two-component current density can be derived in the following manner.
The current density is given by\cite{Faegri07_book}
\begin{align}
    \mathbf{J}(\mathbf{r}) &= \frac{1}{m_e} \mathfrak{R} \left\{ \psi^\dagger \boldsymbol{\sigma} (\boldsymbol{\sigma} \cdot \boldsymbol{\pi}) \psi \right\} \notag
\end{align}
where the mechanical momentum $\boldsymbol{\pi} = \mathbf{p} - q \mathbf{A}$ includes the linear momentum $\mathbf{p} = -i\boldsymbol\nabla$ and the vector potential $\mathbf{A} = \frac{1}{2}\mathbf{B}\times\mathbf{r}$ for electrons and $\varepsilon_{abc}$ is the standard Levi-Civita tensor.
Using the Dirac identity, followed by mathematical derivations and rearrangement (see SI for details), we arrive at the following working expression for computing the \textit{ab initio} current:

\begin{align}
    \mathbf{J}(\mathbf{r}) &= \frac{1}{m_e} \mathfrak{I} \left\{ \psi^\dagger \boldsymbol{\nabla} \psi \right\} - \frac{q}{m_e} \mathfrak{R} \left\{ \psi^\dagger \mathbf{A} \psi \right\} \notag\\
    &+ \frac{1}{2 m_e} \mathfrak{R} \left\{ \left ( \boldsymbol{\nabla} \times (\psi^\dagger \boldsymbol{\sigma} \psi) \right) \right\} 
    \label{eq:gordonx2c}
\end{align}
The three terms in \cref{eq:gordonx2c} are often referred to as the paramagnetic, the diamagnetic, and the spin-dependent paramagnetic contributions to the current density, respectively. Although the last two terms are fully real, we retain the explicit notation to provide clarity when presenting the expressions in the Pauli representation. The programmable expressions in the atomic orbital basis are provided in the SI.

\subsection*{Four-Current-Induced Magnetic Field}

The relativistic magnetic field arising from a four-current density can be obtained from Jefimenko's equation of the magnetic field, \cite{Jefimenko_book, Markus_book}
\begin{align}
    \mathbf{B}(\mathbf{r}, t_r) &= 
    -\frac{\mu_0}{4\pi} 
    \int \int \int
    \Bigg[\frac{(\mathbf{r} - \mathbf{r}')}{|\mathbf{r} - \mathbf{r}'|^3} \times \mathbf{J}(\mathbf{r}', t_r)
    \notag\\
    &+ \frac{(\mathbf{r} - \mathbf{r}')}{|\mathbf{r} - \mathbf{r}'|^2} \times \frac{1}{c}
    \frac{\partial \mathbf{J}(\mathbf{r}', t_r)}{\partial t}\Bigg] d^3\mathbf{r}' \label{eq:jefimenko_b}
\end{align} 
In \cref{eq:jefimenko_b}, the retarded time, $t_r$, is relativistic phenomena arising from the finite speed of light,
\begin{equation}
    t_r = t - \frac{|\mathbf{r} - \mathbf{r}'|}{c}.
\end{equation}
In this work, the relevant velocities are much smaller than the speed of light; therefore, it is safe to neglect relativistic retardation effects. In the physical limit of a slowly varying current density, where its time dependence can be ignored, \cref{eq:jefimenko_b} reduces to the familiar Biot–Savart law.

\subsection*{Computational Methodology}

We constructed a periodic single helicene chain in an orthogonal lattice with lattice constants $a = 20$ \AA~ and $b = 20$ \AA~ to eliminate interchain interactions along the $x$- and $y$-directions. The helical axis is along the $z$-direction. The lattice constant along $z$ was optimized using the Vienna \textit{Ab initio} Simulation Package (VASP).\cite{Furthmuller96_11169} The electron–ion interaction was treated using the projector augmented-wave (PAW) method\cite{Joubert99_1758} with a plane-wave energy cutoff of 525~eV. For the exchange–correlation functional, the Perdew--Burke--Ernzerhof (PBE) form of the generalized gradient approximation (GGA)\cite{Ernzerhof96_3865} was used, along with Grimme’s DFT-D3 dispersion correction.\cite{Krieg10_15} A $1 \times 1 \times 10$ Gamma-centered Monkhorst--Pack $k$- point mesh was employed. The optimized lattice constants were $a = 20 $ \AA, $b = 20$ \AA, $c = 3.515$ \AA~ for both left- and right-handed helicenes.

The helicene molecule was cleaved from the periodic model and further geometry-optimized using the Gaussian~16 package with the B3LYP exchange–correlation functional and Def2-SV(P) basis set. Grimme’s DFT-D3 dispersion correction with Becke--Johnson damping (DFT-D3BJ) was also included.\cite{Goerigk11_1456} The optimized helicene exhibited no imaginary frequencies, confirming a stable minimum.

To investigate the electronic dynamics, we employed the relativistic RT-X2C-TDDFT approach using a development version of the open-source \textsc{Chronus Quantum} package.\cite{Li20_e1436,chronusq_beta2} The initial state ($t=0$) was prepared via a self-consistent field (SCF) calculation under a static electric field of 0.01~au along the $-z$-direction. To initialize the electronic dynamics, an electric field of equal magnitude but opposite direction was applied, and the system was propagated in real time using the RT-X2C-TDDFT method. The time evolution was carried out for 3~fs with a time step of 0.05~au At each step, the time-dependent two-component density was Gordon-decomposed into four-current. All real time calculations used the 6-311G(d,p) basis set.

\section{Code Availability}
The real-time X2C-TDDFT framework is implemented in the public version of the Chronus Quantum software package.\cite{Li20_e1436}  All calculations were performed using executables compiled with GCC version 13.2. 

\section{Data Availability}
All datasets discussed in this work are available from Zenodo (\url{doi.org/10.5281/zenodo.17094822}).

\section{Author Contributions}
S. Upadhyay, X. Zheng, and X. Li conceived the project and developed the computer code.  S. Upadhyay, X. Zheng, T. Wang, and A. Shayit carried out the calculations and analyzed the data. J. Liu, D. Sun, and X. Li acquired research funding. S. Upadhyay, X. Zheng, and X. Li wrote the manuscript with input from all authors.  All authors discussed the results and approved the final version of the manuscript.

\section*{Acknowledgement}
The study of the CISS effect is supported by the Air Force Office of Scientific Research, Multidisciplinary University Research Initiatives (MURI) Program under award number FA9550-23-1-0311. The study of spin-dependent chemical processes is supported by the Air Force Office of Scientific Research (Award No. FA9550-21-1-0344 to XL). The development of relativistic time-dependent method is supported by the Department of Energy in the Computational and Theoretical Chemistry program (Grant No. DE-SC0006863). The Chronus Quantum computational software development
is supported by the Oﬃce of Advanced Cyberinfrastructure, National Science Foundation
(Grant No. OAC-2103717).

\bibliography{References/Journal_Short_Name,References/Zheng_References, References/Li_Group_References.bib,References/x2c_ref,References/Miscellaneous_References,References/All_References,References/TDCC-TDCI,References/4-current}

\providecommand{\latin}[1]{#1}
\makeatletter
\providecommand{\doi}
  {\begingroup\let\do\@makeother\dospecials
  \catcode`\{=1 \catcode`\}=2 \doi@aux}
\providecommand{\doi@aux}[1]{\endgroup\texttt{#1}}
\makeatother
\providecommand*\mcitethebibliography{\thebibliography}
\csname @ifundefined\endcsname{endmcitethebibliography}
  {\let\endmcitethebibliography\endthebibliography}{}
\begin{mcitethebibliography}{161}
\providecommand*\natexlab[1]{#1}
\providecommand*\mciteSetBstSublistMode[1]{}
\providecommand*\mciteSetBstMaxWidthForm[2]{}
\providecommand*\mciteBstWouldAddEndPuncttrue
  {\def\EndOfBibitem{\unskip.}}
\providecommand*\mciteBstWouldAddEndPunctfalse
  {\let\EndOfBibitem\relax}
\providecommand*\mciteSetBstMidEndSepPunct[3]{}
\providecommand*\mciteSetBstSublistLabelBeginEnd[3]{}
\providecommand*\EndOfBibitem{}
\mciteSetBstSublistMode{f}
\mciteSetBstMaxWidthForm{subitem}{(\alph{mcitesubitemcount})}
\mciteSetBstSublistLabelBeginEnd
  {\mcitemaxwidthsubitemform\space}
  {\relax}
  {\relax}

\bibitem[Testa(1986)]{Testa86_60}
Testa,~B. Chiral Aspects of Drug Metabolism. \emph{Trends Pharmacol. Sci}
  \textbf{1986}, 60--64\relax
\mciteBstWouldAddEndPuncttrue
\mciteSetBstMidEndSepPunct{\mcitedefaultmidpunct}
{\mcitedefaultendpunct}{\mcitedefaultseppunct}\relax
\EndOfBibitem
\bibitem[Verbiest \latin{et~al.}(1994)Verbiest, Kauranen, Persoons, Ikonen,
  Kurkela, Lemmetyinen, and Freeman]{Freeman94_9203}
Verbiest,~T.; Kauranen,~M.; Persoons,~A.; Ikonen,~M.; Kurkela,~J.;
  Lemmetyinen,~H.; Freeman,~W.~H. Molecular Optical Activity and the Chiral
  Discriminations. \emph{J. Am. Chem. Soc.} \textbf{1994}, \emph{116},
  9203--9205\relax
\mciteBstWouldAddEndPuncttrue
\mciteSetBstMidEndSepPunct{\mcitedefaultmidpunct}
{\mcitedefaultendpunct}{\mcitedefaultseppunct}\relax
\EndOfBibitem
\bibitem[Krüger and Lö(2000)Krüger, and Lö]{lo20_7031}
Krüger,~P.; Lö,~M. Molecular Chirality and Domain Shapes in Lipid Monolayers
  on Aqueous Surfaces. \emph{Phys. Rev. E} \textbf{2000}, 7031--7043\relax
\mciteBstWouldAddEndPuncttrue
\mciteSetBstMidEndSepPunct{\mcitedefaultmidpunct}
{\mcitedefaultendpunct}{\mcitedefaultseppunct}\relax
\EndOfBibitem
\bibitem[Srinivas \latin{et~al.}(2001)Srinivas, Barbhaiya, and
  Midha]{Midha01_1205}
Srinivas,~N.~R.; Barbhaiya,~R.~H.; Midha,~K.~K. Enantiomeric Drug Development:
  Issues, Considerations, and Regulatory Requirements. \emph{J. Pharm. Sci.}
  \textbf{2001}, 1205--1215\relax
\mciteBstWouldAddEndPuncttrue
\mciteSetBstMidEndSepPunct{\mcitedefaultmidpunct}
{\mcitedefaultendpunct}{\mcitedefaultseppunct}\relax
\EndOfBibitem
\bibitem[Kwiecińska and Cieplak(2005)Kwiecińska, and Cieplak]{Cieplak05_1565}
Kwiecińska,~J.~I.; Cieplak,~M. Chirality and Protein Folding. \emph{J.
  Phys.--Condens. Mat.} \textbf{2005}, \emph{17}, 1565--1580\relax
\mciteBstWouldAddEndPuncttrue
\mciteSetBstMidEndSepPunct{\mcitedefaultmidpunct}
{\mcitedefaultendpunct}{\mcitedefaultseppunct}\relax
\EndOfBibitem
\bibitem[Sandars(2005)]{Sandars05_49}
Sandars,~P.~G. Chirality in the RNA World and Beyond. \emph{Int. J.
  Astrobiology} \textbf{2005}, \emph{4}, 49--61\relax
\mciteBstWouldAddEndPuncttrue
\mciteSetBstMidEndSepPunct{\mcitedefaultmidpunct}
{\mcitedefaultendpunct}{\mcitedefaultseppunct}\relax
\EndOfBibitem
\bibitem[Senge \latin{et~al.}(2014)Senge, Ryan, Letchford, MacGowan, and
  Mielke]{Mielke14_781}
Senge,~M.~O.; Ryan,~A.~A.; Letchford,~K.~A.; MacGowan,~S.~A.; Mielke,~T.
  Chlorophylls, Symmetry, Chirality, and Photosynthesis. \emph{Symmetry}
  \textbf{2014}, \emph{6}, 781--843\relax
\mciteBstWouldAddEndPuncttrue
\mciteSetBstMidEndSepPunct{\mcitedefaultmidpunct}
{\mcitedefaultendpunct}{\mcitedefaultseppunct}\relax
\EndOfBibitem
\bibitem[Luo and Hore(2021)Luo, and Hore]{Hore21_043032}
Luo,~J.; Hore,~P.~J. Chiral-induced Spin Selectivity in the Formation and
  Recombination of Radical Pairs: Cryptochrome Magnetoreception and EPR
  Detection. \emph{New J. Phys.} \textbf{2021}, \emph{23}, 043032\relax
\mciteBstWouldAddEndPuncttrue
\mciteSetBstMidEndSepPunct{\mcitedefaultmidpunct}
{\mcitedefaultendpunct}{\mcitedefaultseppunct}\relax
\EndOfBibitem
\bibitem[Rikken(2011)]{Rikken11_864}
Rikken,~G.~L. A New Twist on Spintronics. \emph{Science} \textbf{2011},
  \emph{331}, 864--865\relax
\mciteBstWouldAddEndPuncttrue
\mciteSetBstMidEndSepPunct{\mcitedefaultmidpunct}
{\mcitedefaultendpunct}{\mcitedefaultseppunct}\relax
\EndOfBibitem
\bibitem[Naaman and Waldeck(2012)Naaman, and Waldeck]{Waldeck12_2178}
Naaman,~R.; Waldeck,~D.~H. Chiral-induced Spin Selectivity Effect. \emph{J.
  Phys. Chem. Lett.} \textbf{2012}, \emph{3}, 2178--2187\relax
\mciteBstWouldAddEndPuncttrue
\mciteSetBstMidEndSepPunct{\mcitedefaultmidpunct}
{\mcitedefaultendpunct}{\mcitedefaultseppunct}\relax
\EndOfBibitem
\bibitem[Guo and Sun(2012)Guo, and Sun]{Sun12_218102}
Guo,~A.~M.; Sun,~Q.~F. Spin-selective Transport of Electrons in DNA Double
  Helix. \emph{Phys. Rev. Lett.} \textbf{2012}, \emph{108}, 218102\relax
\mciteBstWouldAddEndPuncttrue
\mciteSetBstMidEndSepPunct{\mcitedefaultmidpunct}
{\mcitedefaultendpunct}{\mcitedefaultseppunct}\relax
\EndOfBibitem
\bibitem[Gutierrez \latin{et~al.}(2012)Gutierrez, Díaz, Naaman, and
  Cuniberti]{Cuniberti12_081404}
Gutierrez,~R.; Díaz,~E.; Naaman,~R.; Cuniberti,~G. Spin-selective Transport
  Through Helical Molecular Systems. \emph{Phys. Rev. B} \textbf{2012},
  \emph{85}, 081404\relax
\mciteBstWouldAddEndPuncttrue
\mciteSetBstMidEndSepPunct{\mcitedefaultmidpunct}
{\mcitedefaultendpunct}{\mcitedefaultseppunct}\relax
\EndOfBibitem
\bibitem[Eremko and Loktev(2013)Eremko, and Loktev]{Loktev13_165409}
Eremko,~A.~A.; Loktev,~V.~M. Spin Sensitive Electron Transmission Through
  Helical Potentials. \emph{Phys. Rev. B} \textbf{2013}, \emph{88},
  165409\relax
\mciteBstWouldAddEndPuncttrue
\mciteSetBstMidEndSepPunct{\mcitedefaultmidpunct}
{\mcitedefaultendpunct}{\mcitedefaultseppunct}\relax
\EndOfBibitem
\bibitem[Guo and Sun(2014)Guo, and Sun]{Sun14_11658}
Guo,~A.~M.; Sun,~Q.~F. Spin-dependent Electron Transport in Protein-like
  Single-helical Molecules. \emph{Proc. Natl. Acad. Sci. U.S.A.} \textbf{2014},
  \emph{111}, 11658--11662\relax
\mciteBstWouldAddEndPuncttrue
\mciteSetBstMidEndSepPunct{\mcitedefaultmidpunct}
{\mcitedefaultendpunct}{\mcitedefaultseppunct}\relax
\EndOfBibitem
\bibitem[Naaman and Waldeck(2015)Naaman, and Waldeck]{Waldeck15_263}
Naaman,~R.; Waldeck,~D.~H. Spintronics and Chirality: Spin Selectivity in
  Electron Transport Through Chiral Molecules. \emph{Annu. Rev. Phys. Chem.}
  \textbf{2015}, \emph{66}, 263--281\relax
\mciteBstWouldAddEndPuncttrue
\mciteSetBstMidEndSepPunct{\mcitedefaultmidpunct}
{\mcitedefaultendpunct}{\mcitedefaultseppunct}\relax
\EndOfBibitem
\bibitem[Mondal \latin{et~al.}(2016)Mondal, Fontanesi, Waldeck, and
  Naaman]{Naaman16_2560}
Mondal,~P.~C.; Fontanesi,~C.; Waldeck,~D.~H.; Naaman,~R. Spin-Dependent
  Transport through Chiral Molecules Studied by Spin-Dependent
  Electrochemistry. \emph{Acc. Chem. Res.} \textbf{2016}, \emph{49},
  2560--2568\relax
\mciteBstWouldAddEndPuncttrue
\mciteSetBstMidEndSepPunct{\mcitedefaultmidpunct}
{\mcitedefaultendpunct}{\mcitedefaultseppunct}\relax
\EndOfBibitem
\bibitem[Michaeli \latin{et~al.}(2017)Michaeli, Varade, Naaman, and
  Waldeck]{Waldeck17_1}
Michaeli,~K.; Varade,~V.; Naaman,~R.; Waldeck,~D.~H. A New Approach Towards
  Spintronics-spintronics With No Magnets. \emph{J. Phys.--Condens. Mat.}
  \textbf{2017}, \emph{29}, 1--8\relax
\mciteBstWouldAddEndPuncttrue
\mciteSetBstMidEndSepPunct{\mcitedefaultmidpunct}
{\mcitedefaultendpunct}{\mcitedefaultseppunct}\relax
\EndOfBibitem
\bibitem[Naaman \latin{et~al.}(2019)Naaman, Paltiel, and
  Waldeck]{Waldeck19_250}
Naaman,~R.; Paltiel,~Y.; Waldeck,~D.~H. Chiral Molecules and The Electron Spin.
  \emph{Nat. Rev. Chem.} \textbf{2019}, \emph{3}, 250--260\relax
\mciteBstWouldAddEndPuncttrue
\mciteSetBstMidEndSepPunct{\mcitedefaultmidpunct}
{\mcitedefaultendpunct}{\mcitedefaultseppunct}\relax
\EndOfBibitem
\bibitem[Naaman \latin{et~al.}(2019)Naaman, Waldeck, and
  Paltiel]{Paltiel19_133701}
Naaman,~R.; Waldeck,~D.~H.; Paltiel,~Y. Chiral Molecules-ferromagnetic
  Interfaces, an Approach Towards Spin Controlled Interactions. \emph{Appl.
  Phys. Lett.} \textbf{2019}, \emph{115}, 133701\relax
\mciteBstWouldAddEndPuncttrue
\mciteSetBstMidEndSepPunct{\mcitedefaultmidpunct}
{\mcitedefaultendpunct}{\mcitedefaultseppunct}\relax
\EndOfBibitem
\bibitem[Fransson(2019)]{Fransson19_7126}
Fransson,~J. Chirality-Induced Spin Selectivity: The Role of Electron
  Correlations. \emph{J. Phys. Chem. Lett.} \textbf{2019}, \emph{10},
  7126--7132\relax
\mciteBstWouldAddEndPuncttrue
\mciteSetBstMidEndSepPunct{\mcitedefaultmidpunct}
{\mcitedefaultendpunct}{\mcitedefaultseppunct}\relax
\EndOfBibitem
\bibitem[Yu \latin{et~al.}(2022)Yu, Luo, and Bauer]{Bauer22_1}
Yu,~T.; Luo,~Z.; Bauer,~G. E.~W. Chirality as Generalized Spin-Orbit
  Interaction in Spintronics. \emph{arxiv} \textbf{2022}, 1--136\relax
\mciteBstWouldAddEndPuncttrue
\mciteSetBstMidEndSepPunct{\mcitedefaultmidpunct}
{\mcitedefaultendpunct}{\mcitedefaultseppunct}\relax
\EndOfBibitem
\bibitem[Ray \latin{et~al.}(1999)Ray, Ananthavel, Waldeck, and
  Naaman]{Naaman99_814}
Ray,~K.; Ananthavel,~S.~P.; Waldeck,~D.~H.; Naaman,~R. Asymmetric Scattering of
  Polarized Electrons by Organized Organic Films of Chiral Molecules.
  \emph{Science} \textbf{1999}, \emph{283}, 814--816\relax
\mciteBstWouldAddEndPuncttrue
\mciteSetBstMidEndSepPunct{\mcitedefaultmidpunct}
{\mcitedefaultendpunct}{\mcitedefaultseppunct}\relax
\EndOfBibitem
\bibitem[Bloom \latin{et~al.}(2024)Bloom, Paltiel, Naaman, and
  Waldeck]{Waldeck24_1590}
Bloom,~B.~P.; Paltiel,~Y.; Naaman,~R.; Waldeck,~D.~H. Chiral Induced Spin
  Selectivity. \emph{Chem. Rev.} \textbf{2024}, \emph{124}, 1590--1991\relax
\mciteBstWouldAddEndPuncttrue
\mciteSetBstMidEndSepPunct{\mcitedefaultmidpunct}
{\mcitedefaultendpunct}{\mcitedefaultseppunct}\relax
\EndOfBibitem
\bibitem[Hamelbeck \latin{et~al.}(2011)Hamelbeck, Markus, Kettner, Hanne,
  Vager, Naaman, and Zacharias]{Zacharias11_894}
Hamelbeck,~G. B.~V.; Markus,~T.~Z.; Kettner,~M.; Hanne,~G.~F.; Vager,~Z.;
  Naaman,~R.; Zacharias,~H. Spin Selectivity in Electron Transmission Through
  Self-Assembled Monolayers of Double-stranded DNA. \emph{Science}
  \textbf{2011}, \emph{331}, 894--897\relax
\mciteBstWouldAddEndPuncttrue
\mciteSetBstMidEndSepPunct{\mcitedefaultmidpunct}
{\mcitedefaultendpunct}{\mcitedefaultseppunct}\relax
\EndOfBibitem
\bibitem[Xie \latin{et~al.}(2011)Xie, Markus, Cohen, Vager, Gutierrez, and
  Naaman]{Naaman11_4652}
Xie,~Z.; Markus,~T.~Z.; Cohen,~S.~R.; Vager,~Z.; Gutierrez,~R.; Naaman,~R. Spin
  Specific Electron Conduction Through DNA Oligomers. \emph{Nano Lett.}
  \textbf{2011}, \emph{11}, 4652--4655\relax
\mciteBstWouldAddEndPuncttrue
\mciteSetBstMidEndSepPunct{\mcitedefaultmidpunct}
{\mcitedefaultendpunct}{\mcitedefaultseppunct}\relax
\EndOfBibitem
\bibitem[Dhurba and Galperin(2013)Dhurba, and Galperin]{Galperin13_13730}
Dhurba,~R.; Galperin,~M. Electrically Driven Spin Currents in DNA. \emph{Nano
  Lett.} \textbf{2013}, \emph{117}, 13730--13737\relax
\mciteBstWouldAddEndPuncttrue
\mciteSetBstMidEndSepPunct{\mcitedefaultmidpunct}
{\mcitedefaultendpunct}{\mcitedefaultseppunct}\relax
\EndOfBibitem
\bibitem[Kiran \latin{et~al.}(2016)Kiran, Mathew, Cohen, Delgado, Lacour, and
  Naaman]{Naaman16_9203}
Kiran,~V.; Mathew,~S.~P.; Cohen,~S.~R.; Delgado,~I.~H.; Lacour,~J.; Naaman,~R.
  Helicenes - A New Class of Organic Spin Filter. \emph{Adv. Mater.}
  \textbf{2016}, \emph{28}, 1957--1962\relax
\mciteBstWouldAddEndPuncttrue
\mciteSetBstMidEndSepPunct{\mcitedefaultmidpunct}
{\mcitedefaultendpunct}{\mcitedefaultseppunct}\relax
\EndOfBibitem
\bibitem[Kettner \latin{et~al.}(2018)Kettner, Maslyuk, Nürenberg, Seibel,
  Gutierrez, Cuniberti, Ernst, and Zacharias]{Zacharias18_2025}
Kettner,~M.; Maslyuk,~V.~V.; Nürenberg,~D.; Seibel,~J.; Gutierrez,~R.;
  Cuniberti,~G.; Ernst,~K.~H.; Zacharias,~H. Chirality-Dependent Electron Spin
  Filtering by Molecular Monolayers of Helicenes. \emph{J. Phys. Chem. Lett.}
  \textbf{2018}, \emph{9}, 2025--2030\relax
\mciteBstWouldAddEndPuncttrue
\mciteSetBstMidEndSepPunct{\mcitedefaultmidpunct}
{\mcitedefaultendpunct}{\mcitedefaultseppunct}\relax
\EndOfBibitem
\bibitem[Safari \latin{et~al.}(2022)Safari, Matthes, Ernst, Bürgler, and
  Schneider]{Schneider22_1}
Safari,~M.~R.; Matthes,~F.; Ernst,~K.~H.; Bürgler,~D.~E.; Schneider,~C.~M.
  Deposition of Chiral Heptahelicene Molecules on Ferromagnetic Co and Fe
  Thin-Film Substrates. \emph{Nanomater.} \textbf{2022}, \emph{12}, 1--18\relax
\mciteBstWouldAddEndPuncttrue
\mciteSetBstMidEndSepPunct{\mcitedefaultmidpunct}
{\mcitedefaultendpunct}{\mcitedefaultseppunct}\relax
\EndOfBibitem
\bibitem[Liang \latin{et~al.}(2022)Liang, Banjac, Martin, Zigon, Lee,
  Vanthuyne, Garcés-Pineda, Galán-Mascarós, Hu, Avarvari, and
  Lingenfelder]{Lingenfelder22_1}
Liang,~Y.; Banjac,~K.; Martin,~K.; Zigon,~N.; Lee,~S.; Vanthuyne,~N.;
  Garcés-Pineda,~F.~A.; Galán-Mascarós,~J.~R.; Hu,~X.; Avarvari,~N.;
  Lingenfelder,~M. Enhancement of electrocatalytic oxygen evolution by chiral
  molecular functionalization of hybrid 2D electrodes. \emph{Nat. Chem.}
  \textbf{2022}, \emph{13}, 1--9\relax
\mciteBstWouldAddEndPuncttrue
\mciteSetBstMidEndSepPunct{\mcitedefaultmidpunct}
{\mcitedefaultendpunct}{\mcitedefaultseppunct}\relax
\EndOfBibitem
\bibitem[Rodríguez \latin{et~al.}(2022)Rodríguez, Naranjo, Kumar, Matozzo,
  Das, Zhu, Vanthuyne, Gómez, Naaman, Sánchez, and Crassous]{Crassous22_7709}
Rodríguez,~R.; Naranjo,~C.; Kumar,~A.; Matozzo,~P.; Das,~T.~K.; Zhu,~Q.;
  Vanthuyne,~N.; Gómez,~R.; Naaman,~R.; Sánchez,~L.; Crassous,~J. Mutual
  Monomer Orientation to Bias the Supramolecular Polymerization of [6]Helicenes
  and the Resulting Circularly Polarized Light and Spin Filtering Properties.
  \emph{J. Am. Chem. Soc.} \textbf{2022}, \emph{144}, 7709--7719\relax
\mciteBstWouldAddEndPuncttrue
\mciteSetBstMidEndSepPunct{\mcitedefaultmidpunct}
{\mcitedefaultendpunct}{\mcitedefaultseppunct}\relax
\EndOfBibitem
\bibitem[Rodríguez \latin{et~al.}(2023)Rodríguez, Naranjo, Kumar, Dhbaibi,
  Matozzo, Camerel, Vanthuyne, Gómez, Naaman, Sánchez, and
  Crassous]{Crassous23_1}
Rodríguez,~R.; Naranjo,~C.; Kumar,~A.; Dhbaibi,~K.; Matozzo,~P.; Camerel,~F.;
  Vanthuyne,~N.; Gómez,~R.; Naaman,~R.; Sánchez,~L.; Crassous,~J. Weakly
  Self-Assembled [6]Helicenes: Circularly Polarized Light and Spin Filtering
  Properties. \emph{Chem. Eur. J.} \textbf{2023}, \emph{29}, 1--7\relax
\mciteBstWouldAddEndPuncttrue
\mciteSetBstMidEndSepPunct{\mcitedefaultmidpunct}
{\mcitedefaultendpunct}{\mcitedefaultseppunct}\relax
\EndOfBibitem
\bibitem[Giaconi \latin{et~al.}(2023)Giaconi, Poggini, Lupi, Briganti, Kumar,
  Das, Sorrentino, Viglianisi, Menichetti, Naaman, Sessoli, and
  Mannini]{Mannini23_15189}
Giaconi,~N.; Poggini,~L.; Lupi,~M.; Briganti,~M.; Kumar,~A.; Das,~T.~K.;
  Sorrentino,~A.~L.; Viglianisi,~C.; Menichetti,~S.; Naaman,~R.; Sessoli,~R.;
  Mannini,~M. Efficient Spin-Selective Electron Transport at Low Voltages of
  Thia-Bridged Triarylamine Hetero[4]helicenes Chemisorbed Monolayer. \emph{ACS
  Nano} \textbf{2023}, \emph{17}, 15189--15198\relax
\mciteBstWouldAddEndPuncttrue
\mciteSetBstMidEndSepPunct{\mcitedefaultmidpunct}
{\mcitedefaultendpunct}{\mcitedefaultseppunct}\relax
\EndOfBibitem
\bibitem[Safari \latin{et~al.}(2024)Safari, Matthes, Schneider, Ernst, and
  Bürgler]{Burgler24_2308233}
Safari,~M.~R.; Matthes,~F.; Schneider,~C.~M.; Ernst,~K.~H.; Bürgler,~D.~E.
  Spin-Selective Electron Transport Through Single Chiral Molecules.
  \emph{Small} \textbf{2024}, \emph{20}, 2308233\relax
\mciteBstWouldAddEndPuncttrue
\mciteSetBstMidEndSepPunct{\mcitedefaultmidpunct}
{\mcitedefaultendpunct}{\mcitedefaultseppunct}\relax
\EndOfBibitem
\bibitem[Mishra \latin{et~al.}(2013)Mishra, Markus, Naaman, Kettner, Göhler,
  Zacharias, Friedman, Sheves, and Fontanesi]{Fontanesi13_14872}
Mishra,~D.; Markus,~T.~Z.; Naaman,~R.; Kettner,~M.; Göhler,~B.; Zacharias,~H.;
  Friedman,~N.; Sheves,~M.; Fontanesi,~C. Spin-dependent Electron Transmission
  Through Bacteriorhodopsin Embedded in Purple Membrane. \emph{Proc. Natl.
  Acad. Sci. U.S.A.} \textbf{2013}, \emph{110}, 14872--14876\relax
\mciteBstWouldAddEndPuncttrue
\mciteSetBstMidEndSepPunct{\mcitedefaultmidpunct}
{\mcitedefaultendpunct}{\mcitedefaultseppunct}\relax
\EndOfBibitem
\bibitem[Carmeli \latin{et~al.}(2014)Carmeli, Kumar, Heifler, Carmeli, and
  Naaman]{Naaman14_8953}
Carmeli,~I.; Kumar,~K.~S.; Heifler,~O.; Carmeli,~C.; Naaman,~R. Spin
  Selectivity in Electron Transfer in Photosystem I. \emph{Angew. Chem.}
  \textbf{2014}, \emph{53}, 8953--8958\relax
\mciteBstWouldAddEndPuncttrue
\mciteSetBstMidEndSepPunct{\mcitedefaultmidpunct}
{\mcitedefaultendpunct}{\mcitedefaultseppunct}\relax
\EndOfBibitem
\bibitem[Mondal \latin{et~al.}(2015)Mondal, Fontanesi, Waldeck, and
  Naaman]{Naaman15_3377}
Mondal,~P.~C.; Fontanesi,~C.; Waldeck,~D.~H.; Naaman,~R. Field and Chirality
  Effects on Electrochemical Charge Transfer Rates: Spin Dependent
  Electrochemistry. \emph{ACS Nano} \textbf{2015}, \emph{9}, 3377--3384\relax
\mciteBstWouldAddEndPuncttrue
\mciteSetBstMidEndSepPunct{\mcitedefaultmidpunct}
{\mcitedefaultendpunct}{\mcitedefaultseppunct}\relax
\EndOfBibitem
\bibitem[Varade \latin{et~al.}(2018)Varade, Markus, Vankayala, Friedman,
  Sheves, Waldeck, and Naaman]{Naaman18_1091}
Varade,~V.; Markus,~T.; Vankayala,~K.; Friedman,~N.; Sheves,~M.;
  Waldeck,~D.~H.; Naaman,~R. Bacteriorhodopsin Based Non-Magnetic Spin Filters
  for Biomolecular Spintronics. \emph{Phys. Chem. Chem. Phys.} \textbf{2018},
  \emph{20}, 1091--1097\relax
\mciteBstWouldAddEndPuncttrue
\mciteSetBstMidEndSepPunct{\mcitedefaultmidpunct}
{\mcitedefaultendpunct}{\mcitedefaultseppunct}\relax
\EndOfBibitem
\bibitem[Mishra \latin{et~al.}(2019)Mishra, Pirbadian, Mondal, El-Naggar, and
  Naaman]{Naaman19_19198}
Mishra,~S.; Pirbadian,~S.; Mondal,~A.~K.; El-Naggar,~M.~Y.; Naaman,~R.
  Spin-Dependent Electron Transport Through Bacterial Cell Surface Multiheme
  Electron Conduits. \emph{J. Am. Chem. Soc.} \textbf{2019}, \emph{141},
  19198--19202\relax
\mciteBstWouldAddEndPuncttrue
\mciteSetBstMidEndSepPunct{\mcitedefaultmidpunct}
{\mcitedefaultendpunct}{\mcitedefaultseppunct}\relax
\EndOfBibitem
\bibitem[Kashiwagi \latin{et~al.}(2020)Kashiwagi, Tassinari, Haraguchi,
  Banerjee-Gosh, Akitsu, and Naaman]{Naaman20_808}
Kashiwagi,~K.; Tassinari,~F.; Haraguchi,~T.; Banerjee-Gosh,~K.; Akitsu,~T.;
  Naaman,~R. Electron Transfer Via Helical Oligopeptide to Laccase Including
  Chiral Schiff Base Copper Mediators. \emph{Symmetry} \textbf{2020},
  \emph{12}, 808\relax
\mciteBstWouldAddEndPuncttrue
\mciteSetBstMidEndSepPunct{\mcitedefaultmidpunct}
{\mcitedefaultendpunct}{\mcitedefaultseppunct}\relax
\EndOfBibitem
\bibitem[Banerjee-Ghosh \latin{et~al.}(2020)Banerjee-Ghosh, Ghosh, Mazal,
  Riven, Haran, and Naaman]{Naaman20_20456}
Banerjee-Ghosh,~K.; Ghosh,~S.; Mazal,~H.; Riven,~I.; Haran,~G.; Naaman,~R.
  Long-Range Charge Reorganization as an Allosteric Control Signal in Proteins.
  \emph{J. Am. Chem. Soc.} \textbf{2020}, \emph{142}, 20456--20462\relax
\mciteBstWouldAddEndPuncttrue
\mciteSetBstMidEndSepPunct{\mcitedefaultmidpunct}
{\mcitedefaultendpunct}{\mcitedefaultseppunct}\relax
\EndOfBibitem
\bibitem[Jia \latin{et~al.}(2020)Jia, Wang, Zhang, Yang, and Yan]{Yan20_6607}
Jia,~L.; Wang,~C.; Zhang,~Y.; Yang,~L.; Yan,~Y. Efficient Spin Selectivity in
  Self-Assembled Superhelical Conducting Polymer Microfibers. \emph{ACS Nano}
  \textbf{2020}, \emph{14}, 6607--6615\relax
\mciteBstWouldAddEndPuncttrue
\mciteSetBstMidEndSepPunct{\mcitedefaultmidpunct}
{\mcitedefaultendpunct}{\mcitedefaultseppunct}\relax
\EndOfBibitem
\bibitem[Sang \latin{et~al.}(2021)Sang, Mishra, Tassinari, Karuppannan,
  Carmieli, Teo, Migliore, Beratan, Gray, Pecht, Fransson, Waldeck, and
  Naaman]{Naaman21_9875}
Sang,~Y.; Mishra,~S.; Tassinari,~F.; Karuppannan,~S.~K.; Carmieli,~R.;
  Teo,~R.~D.; Migliore,~A.; Beratan,~D.~N.; Gray,~H.~B.; Pecht,~I.;
  Fransson,~J.; Waldeck,~D.~H.; Naaman,~R. Temperature Dependence of Charge and
  Spin Transfer in Azurin. \emph{J. Phys. Chem. C} \textbf{2021}, \emph{125},
  9875--9883\relax
\mciteBstWouldAddEndPuncttrue
\mciteSetBstMidEndSepPunct{\mcitedefaultmidpunct}
{\mcitedefaultendpunct}{\mcitedefaultseppunct}\relax
\EndOfBibitem
\bibitem[Niman \latin{et~al.}(2023)Niman, Sukenik, Dang, Nwachukwu,
  Thirumurthy, Jones, Naaman, Santra, Das, Paltiel, Baczewski, and
  El-Naggar]{Naggar23_145101}
Niman,~C.~M.; Sukenik,~N.; Dang,~T.; Nwachukwu,~J.; Thirumurthy,~M.~A.;
  Jones,~A.~K.; Naaman,~R.; Santra,~K.; Das,~T.~K.; Paltiel,~Y.;
  Baczewski,~L.~T.; El-Naggar,~M.~Y. Bacterial Extracellular Electron Transfer
  Components are Spin Selective. \emph{J. Chem. Phys.} \textbf{2023},
  \emph{159}, 145101\relax
\mciteBstWouldAddEndPuncttrue
\mciteSetBstMidEndSepPunct{\mcitedefaultmidpunct}
{\mcitedefaultendpunct}{\mcitedefaultseppunct}\relax
\EndOfBibitem
\bibitem[Wei \latin{et~al.}(2023)Wei, Bloom, Dunlap-Shohl, Clever, Rivas, and
  Waldeck]{Waldeck23_6462}
Wei,~J.; Bloom,~B.~P.; Dunlap-Shohl,~W.~A.; Clever,~C.~B.; Rivas,~J.~E.;
  Waldeck,~D.~H. Examining the Effects of Homochirality for Electron Transfer
  in Protein Assemblies. \emph{J. Phys. Chem. B} \textbf{2023}, \emph{127},
  6462--6469\relax
\mciteBstWouldAddEndPuncttrue
\mciteSetBstMidEndSepPunct{\mcitedefaultmidpunct}
{\mcitedefaultendpunct}{\mcitedefaultseppunct}\relax
\EndOfBibitem
\bibitem[Gupta \latin{et~al.}(2023)Gupta, Chinnasamy, Sahu, Matheshwaran, Sow,
  and Mondal]{Mondal23_024708}
Gupta,~R.; Chinnasamy,~H.~V.; Sahu,~D.; Matheshwaran,~S.; Sow,~C.;
  Mondal,~P.~C. Spin-dependent Electrified Protein Interfaces for Probing the
  CISS Effect. \emph{J. Chem. Phys.} \textbf{2023}, \emph{159}, 024708\relax
\mciteBstWouldAddEndPuncttrue
\mciteSetBstMidEndSepPunct{\mcitedefaultmidpunct}
{\mcitedefaultendpunct}{\mcitedefaultseppunct}\relax
\EndOfBibitem
\bibitem[Mondal \latin{et~al.}(2015)Mondal, Kantor-Uriel, Mathew, Tassinari,
  Fontanesi, and Naaman]{Naaman15_1924}
Mondal,~P.~C.; Kantor-Uriel,~N.; Mathew,~S.~P.; Tassinari,~F.; Fontanesi,~C.;
  Naaman,~R. Chiral Conductive Polymers as Spin Filters. \emph{Adv. Mater.}
  \textbf{2015}, \emph{27}, 1924--1927\relax
\mciteBstWouldAddEndPuncttrue
\mciteSetBstMidEndSepPunct{\mcitedefaultmidpunct}
{\mcitedefaultendpunct}{\mcitedefaultseppunct}\relax
\EndOfBibitem
\bibitem[Tassinari \latin{et~al.}(2017)Tassinari, Banerjee-Ghosh, Parenti,
  Kiran, Mucci, and Naaman]{Naaman17_15777}
Tassinari,~F.; Banerjee-Ghosh,~K.; Parenti,~F.; Kiran,~V.; Mucci,~A.;
  Naaman,~R. Enhanced Hydrogen Production with Chiral Conductive Polymer-Based
  Electrodes. \emph{J. Phys. Chem. C} \textbf{2017}, \emph{121},
  15777--15783\relax
\mciteBstWouldAddEndPuncttrue
\mciteSetBstMidEndSepPunct{\mcitedefaultmidpunct}
{\mcitedefaultendpunct}{\mcitedefaultseppunct}\relax
\EndOfBibitem
\bibitem[Mishra \latin{et~al.}(2020)Mishra, Mondal, Smolinsky, Naaman, Maeda,
  Nishimura, Taniguchi, Yoshida, Takayama, and Yashima]{Yashima20_15777}
Mishra,~S.; Mondal,~A.~K.; Smolinsky,~E.~Z.; Naaman,~R.; Maeda,~K.;
  Nishimura,~T.; Taniguchi,~T.; Yoshida,~T.; Takayama,~K.; Yashima,~E. Spin
  Filtering Along Chiral Polymers. \emph{Angew. Chem. Int. Ed.} \textbf{2020},
  \emph{59}, 14671--14676\relax
\mciteBstWouldAddEndPuncttrue
\mciteSetBstMidEndSepPunct{\mcitedefaultmidpunct}
{\mcitedefaultendpunct}{\mcitedefaultseppunct}\relax
\EndOfBibitem
\bibitem[Mondal \latin{et~al.}(2021)Mondal, Preuss, Śleczkowski, Das,
  Vantomme, Meijer, and Naaman]{Naaman21_7189}
Mondal,~A.~K.; Preuss,~M.~D.; Śleczkowski,~M.~L.; Das,~T.~K.; Vantomme,~G.;
  Meijer,~E.~W.; Naaman,~R. Spin Filtering in Supramolecular Polymers Assembled
  from Achiral Monomers Mediated by Chiral Solvents. \emph{J. Am. Chem. Soc.}
  \textbf{2021}, \emph{143}, 7189--7195\relax
\mciteBstWouldAddEndPuncttrue
\mciteSetBstMidEndSepPunct{\mcitedefaultmidpunct}
{\mcitedefaultendpunct}{\mcitedefaultseppunct}\relax
\EndOfBibitem
\bibitem[Bhowmick \latin{et~al.}(2022)Bhowmick, Das, Santra, Mondal, Tassinari,
  Schwarz, Diesendruck, and Naaman]{Naaman22_2727}
Bhowmick,~D.~K.; Das,~T.~K.; Santra,~K.; Mondal,~A.~K.; Tassinari,~F.;
  Schwarz,~R.; Diesendruck,~C.~E.; Naaman,~R. Spin-induced Asymmetry
  Reaction-The formation of Asymmetric Carbon by Electropolymerization.
  \emph{Sci. Adv.} \textbf{2022}, \emph{8}, 2727\relax
\mciteBstWouldAddEndPuncttrue
\mciteSetBstMidEndSepPunct{\mcitedefaultmidpunct}
{\mcitedefaultendpunct}{\mcitedefaultseppunct}\relax
\EndOfBibitem
\bibitem[Nguyen \latin{et~al.}(2023)Nguyen, Paltiel, Baczewski, and
  Tegenkamp]{Tegenkamp23_17406}
Nguyen,~T. N.~H.; Paltiel,~Y.; Baczewski,~L.~T.; Tegenkamp,~C. Spin
  Polarization of Polyalanine Molecules in 2D and Dimer-Row Assemblies Adsorbed
  on Magnetic Substrates: The Role of Coupling, Chirality, and Coordination.
  \emph{ACS Appl. Mater. Interfaces} \textbf{2023}, \emph{15},
  17406--17412\relax
\mciteBstWouldAddEndPuncttrue
\mciteSetBstMidEndSepPunct{\mcitedefaultmidpunct}
{\mcitedefaultendpunct}{\mcitedefaultseppunct}\relax
\EndOfBibitem
\bibitem[Medina \latin{et~al.}(2015)Medina, González-Arraga,
  Finkelstein-Shapiro, Berche, and Mujica]{Mujica15_218102}
Medina,~E.; González-Arraga,~L.~A.; Finkelstein-Shapiro,~D.; Berche,~B.;
  Mujica,~V. Continuum Model for Chiral Induced Spin Selectivity in Helical
  Molecules. \emph{J. Chem. Phys.} \textbf{2015}, \emph{142}, 194308\relax
\mciteBstWouldAddEndPuncttrue
\mciteSetBstMidEndSepPunct{\mcitedefaultmidpunct}
{\mcitedefaultendpunct}{\mcitedefaultseppunct}\relax
\EndOfBibitem
\bibitem[Varela \latin{et~al.}(2016)Varela, Mujica, and
  Medina]{Medina16_155436}
Varela,~S.; Mujica,~V.; Medina,~E. Effective Spin-orbit Couplings in an
  Analytical Tight-binding Model of DNA: Spin Filtering and Chiral Spin
  Transport. \emph{Phys. Rev. B} \textbf{2016}, \emph{93}, 155436\relax
\mciteBstWouldAddEndPuncttrue
\mciteSetBstMidEndSepPunct{\mcitedefaultmidpunct}
{\mcitedefaultendpunct}{\mcitedefaultseppunct}\relax
\EndOfBibitem
\bibitem[Chang \latin{et~al.}(2018)Chang, Wieder, Schindler, Sanchez,
  Belopolski, Huang, Singh, Wu, Chang, Neupert, Xu, Lin, and
  Hasan]{Hasan18_978}
Chang,~G.; Wieder,~B.~J.; Schindler,~F.; Sanchez,~D.~S.; Belopolski,~I.;
  Huang,~S.~M.; Singh,~B.; Wu,~D.; Chang,~T.~R.; Neupert,~T.; Xu,~S.~Y.;
  Lin,~H.; Hasan,~M.~Z. Topological Quantum Properties of Chiral Crystals.
  \emph{Nat. Mater.} \textbf{2018}, \emph{17}, 978--985\relax
\mciteBstWouldAddEndPuncttrue
\mciteSetBstMidEndSepPunct{\mcitedefaultmidpunct}
{\mcitedefaultendpunct}{\mcitedefaultseppunct}\relax
\EndOfBibitem
\bibitem[Dalum and Hedegård(2019)Dalum, and Hedegård]{Hedegard19_5253}
Dalum,~S.; Hedegård,~P. Theory of Chiral Induced Spin Selectivity. \emph{Nano
  Lett.} \textbf{2019}, \emph{19}, 5253--5259\relax
\mciteBstWouldAddEndPuncttrue
\mciteSetBstMidEndSepPunct{\mcitedefaultmidpunct}
{\mcitedefaultendpunct}{\mcitedefaultseppunct}\relax
\EndOfBibitem
\bibitem[Yang \latin{et~al.}(2019)Yang, Wal, and Wees]{Wees19_024418}
Yang,~X.; Wal,~C. H. V.~D.; Wees,~B. J.~V. Spin-dependent Electron Transmission
  Model for Chiral Molecules in Mesoscopic Devices. \emph{Phys. Rev. B}
  \textbf{2019}, \emph{99}, 024418\relax
\mciteBstWouldAddEndPuncttrue
\mciteSetBstMidEndSepPunct{\mcitedefaultmidpunct}
{\mcitedefaultendpunct}{\mcitedefaultseppunct}\relax
\EndOfBibitem
\bibitem[Liu \latin{et~al.}(2021)Liu, Xiao, Koo, and Yan]{Yan21_638}
Liu,~Y.; Xiao,~J.; Koo,~J.; Yan,~B. Chirality-driven Topological Electronic
  Structure of DNA-like Materials. \emph{Nat. Mater.} \textbf{2021}, \emph{20},
  638--644\relax
\mciteBstWouldAddEndPuncttrue
\mciteSetBstMidEndSepPunct{\mcitedefaultmidpunct}
{\mcitedefaultendpunct}{\mcitedefaultseppunct}\relax
\EndOfBibitem
\bibitem[Evers \latin{et~al.}(2022)Evers, Aharony, Bar-Gill, Entin-Wohlman,
  Hedegård, Hod, Jelinek, Kamieniarz, Lemeshko, Michaeli, Mujica, Naaman,
  Paltiel, Refaely-Abramson, Tal, Thijssen, Thoss, van Ruitenbeek,
  Venkataraman, Waldeck, Yan, and Kronik]{Kronik22_2106629}
Evers,~F.; Aharony,~A.; Bar-Gill,~N.; Entin-Wohlman,~O.; Hedegård,~P.;
  Hod,~O.; Jelinek,~P.; Kamieniarz,~G.; Lemeshko,~M.; Michaeli,~K.; Mujica,~V.;
  Naaman,~R.; Paltiel,~Y.; Refaely-Abramson,~S.; Tal,~O.; Thijssen,~J.;
  Thoss,~M.; van Ruitenbeek,~J.~M.; Venkataraman,~L.; Waldeck,~D.~H.; Yan,~B.;
  Kronik,~L. Theory of Chirality Induced Spin Selectivity: Progress and
  Challenges. \emph{Adv. Mater.} \textbf{2022}, \emph{34}, 2106629\relax
\mciteBstWouldAddEndPuncttrue
\mciteSetBstMidEndSepPunct{\mcitedefaultmidpunct}
{\mcitedefaultendpunct}{\mcitedefaultseppunct}\relax
\EndOfBibitem
\bibitem[Huisman \latin{et~al.}(2023)Huisman, Heinisch, and
  Thijssen]{Thijssen23_6900}
Huisman,~K.~H.; Heinisch,~J. B. M.~Y.; Thijssen,~J.~M. Chirality-Induced Spin
  Selectivity (CISS) Effect: Magnetocurrent-Voltage Characteristics with
  Coulomb Interactions I. \emph{J. Phys. Chem. C} \textbf{2023}, \emph{127},
  6900--6905\relax
\mciteBstWouldAddEndPuncttrue
\mciteSetBstMidEndSepPunct{\mcitedefaultmidpunct}
{\mcitedefaultendpunct}{\mcitedefaultseppunct}\relax
\EndOfBibitem
\bibitem[Boulougouris(2013)]{Boulougouris13_114111}
Boulougouris,~G.~C. Multidimensional Direct Free Energy Perturbation. \emph{J.
  Chem. Phys.} \textbf{2013}, \emph{138}, 114111\relax
\mciteBstWouldAddEndPuncttrue
\mciteSetBstMidEndSepPunct{\mcitedefaultmidpunct}
{\mcitedefaultendpunct}{\mcitedefaultseppunct}\relax
\EndOfBibitem
\bibitem[Dubi(2022)]{Dubi22_10878}
Dubi,~Y. Spinterface Chirality-Induced Spin Selectivity Effect in
  Bio-molecules. \emph{Chem. Sci.} \textbf{2022}, \emph{13}, 10878--10883\relax
\mciteBstWouldAddEndPuncttrue
\mciteSetBstMidEndSepPunct{\mcitedefaultmidpunct}
{\mcitedefaultendpunct}{\mcitedefaultseppunct}\relax
\EndOfBibitem
\bibitem[Zöllner \latin{et~al.}(2020)Zöllner, Saghatchi, Mujica, and
  Herrmann]{Herrmann20_7357}
Zöllner,~M.~S.; Saghatchi,~A.; Mujica,~V.; Herrmann,~C. Influence of
  Electronic Structure Modeling and Junction Structure on First-Principles
  Chiral Induced Spin Selectivity. \emph{J. Chem. Theory Comput.}
  \textbf{2020}, \emph{16}, 7357--7371\relax
\mciteBstWouldAddEndPuncttrue
\mciteSetBstMidEndSepPunct{\mcitedefaultmidpunct}
{\mcitedefaultendpunct}{\mcitedefaultseppunct}\relax
\EndOfBibitem
\bibitem[Eckvahl \latin{et~al.}(2023)Eckvahl, Tcyrulnikov, Chiesa, Bradley,
  Young, Carretta, Krzyaniak, and Wasielewski]{Wasielewski23_197}
Eckvahl,~H.~J.; Tcyrulnikov,~N.~A.; Chiesa,~A.; Bradley,~J.~M.; Young,~R.~M.;
  Carretta,~S.; Krzyaniak,~M.~D.; Wasielewski,~M.~R. Direct Observation of
  Chirality-induced Spin Selectivity in Electron Donor-acceptor Molecules.
  \emph{Science} \textbf{2023}, \emph{382}, 197--210\relax
\mciteBstWouldAddEndPuncttrue
\mciteSetBstMidEndSepPunct{\mcitedefaultmidpunct}
{\mcitedefaultendpunct}{\mcitedefaultseppunct}\relax
\EndOfBibitem
\bibitem[Li \latin{et~al.}(2005)Li, Tully, Schlegel, and Frisch]{Li05_084106}
Li,~X.; Tully,~J.~C.; Schlegel,~H.~B.; Frisch,~M.~J. Ab Initio Ehrenfest
  Dynamics. \emph{J. Chem. Phys.} \textbf{2005}, \emph{123}, 084106\relax
\mciteBstWouldAddEndPuncttrue
\mciteSetBstMidEndSepPunct{\mcitedefaultmidpunct}
{\mcitedefaultendpunct}{\mcitedefaultseppunct}\relax
\EndOfBibitem
\bibitem[Castro \latin{et~al.}(2006)Castro, Appel, Oliveira, Rozzi, Andrade,
  Lorenzen, Marques, Gross, and Rubio]{Rubio06_2465}
Castro,~A.; Appel,~H.; Oliveira,~M.; Rozzi,~C.~A.; Andrade,~X.; Lorenzen,~F.;
  Marques,~M.~A.; Gross,~E.~K.; Rubio,~A. Octopus: A Tool for the Application
  of Time-dependent Density Functional Theory. \emph{Phys. Status Solidi B}
  \textbf{2006}, \emph{243}, 2465--2488\relax
\mciteBstWouldAddEndPuncttrue
\mciteSetBstMidEndSepPunct{\mcitedefaultmidpunct}
{\mcitedefaultendpunct}{\mcitedefaultseppunct}\relax
\EndOfBibitem
\bibitem[Isborn \latin{et~al.}(2007)Isborn, Li, and Tully]{Li07_134307}
Isborn,~C.~M.; Li,~X.; Tully,~J.~C. TDDFT Ehrenfest Dynamics: Collisions
  between Atomic Oxygen and Graphite Clusters. \emph{J. Chem. Phys.}
  \textbf{2007}, \emph{126}, 134307\relax
\mciteBstWouldAddEndPuncttrue
\mciteSetBstMidEndSepPunct{\mcitedefaultmidpunct}
{\mcitedefaultendpunct}{\mcitedefaultseppunct}\relax
\EndOfBibitem
\bibitem[Liang \latin{et~al.}(2011)Liang, Chapman, and Li]{Li11_184102}
Liang,~W.; Chapman,~C.~T.; Li,~X. Efficient First-Principles Electronic
  Dynamics. \emph{J. Chem. Phys.} \textbf{2011}, \emph{134}, 184102\relax
\mciteBstWouldAddEndPuncttrue
\mciteSetBstMidEndSepPunct{\mcitedefaultmidpunct}
{\mcitedefaultendpunct}{\mcitedefaultseppunct}\relax
\EndOfBibitem
\bibitem[{DePrince III} \latin{et~al.}(2011){DePrince III}, Pelton, Guest, and
  Gray]{Gray11_196806}
{DePrince III},~A.~E.; Pelton,~M.; Guest,~J.~R.; Gray,~S.~K. Emergence of
  Excited-State Plasmon Modes in Linear Hydrogen Chains from Time-Dependent
  Quantum Mechanical Methods. \emph{Phys. Rev. Lett.} \textbf{2011},
  \emph{107}, 196806\relax
\mciteBstWouldAddEndPuncttrue
\mciteSetBstMidEndSepPunct{\mcitedefaultmidpunct}
{\mcitedefaultendpunct}{\mcitedefaultseppunct}\relax
\EndOfBibitem
\bibitem[Lopata and Govind(2011)Lopata, and Govind]{Govind11_1344}
Lopata,~K.; Govind,~N. Modeling Fast Electron Dynamics with Real-time
  Time-dependent Density Functional Theory: Application to Small Molecules and
  Chromophores. \emph{J. Chem. Theory Comput.} \textbf{2011}, \emph{7},
  1344--1355\relax
\mciteBstWouldAddEndPuncttrue
\mciteSetBstMidEndSepPunct{\mcitedefaultmidpunct}
{\mcitedefaultendpunct}{\mcitedefaultseppunct}\relax
\EndOfBibitem
\bibitem[Lopata \latin{et~al.}(2012)Lopata, Van~Kuiken, Khalil, and
  Govind]{Govind12_3284}
Lopata,~K.; Van~Kuiken,~B.~E.; Khalil,~M.; Govind,~N. {Linear-Response and
  Real-Time Time-Dependent Density Functional Theory Studies of Core-Level
  Near-Edge X-Ray Absorption}. \emph{J. Chem. Theory Comput.} \textbf{2012},
  \emph{8}, 3284--3292\relax
\mciteBstWouldAddEndPuncttrue
\mciteSetBstMidEndSepPunct{\mcitedefaultmidpunct}
{\mcitedefaultendpunct}{\mcitedefaultseppunct}\relax
\EndOfBibitem
\bibitem[Ding \latin{et~al.}(2014)Ding, Goings, Frisch, and Li]{Li14_214111}
Ding,~F.; Goings,~J.~J.; Frisch,~M.~J.; Li,~X. Ab Initio Non-Relativistic Spin
  Dynamics. \emph{J. Chem. Phys.} \textbf{2014}, \emph{141}, 214111\relax
\mciteBstWouldAddEndPuncttrue
\mciteSetBstMidEndSepPunct{\mcitedefaultmidpunct}
{\mcitedefaultendpunct}{\mcitedefaultseppunct}\relax
\EndOfBibitem
\bibitem[Provorse \latin{et~al.}(2015)Provorse, Habenicht, and
  Isborn]{Isborn15_4791}
Provorse,~M.~R.; Habenicht,~B.~F.; Isborn,~C.~M. {Peak-Shifting in Real-Time
  Time-Dependent Density Functional Theory}. \emph{J. Chem. Theory Comput.}
  \textbf{2015}, \emph{11}, 4791--4802\relax
\mciteBstWouldAddEndPuncttrue
\mciteSetBstMidEndSepPunct{\mcitedefaultmidpunct}
{\mcitedefaultendpunct}{\mcitedefaultseppunct}\relax
\EndOfBibitem
\bibitem[Goings and Li(2016)Goings, and Li]{Li16_234102}
Goings,~J.~J.; Li,~X. An Atomic Orbital Based Real-Time Time-Dependent Density
  Functional Theory for Computing Electronic Circular Dichroism Band Spectra.
  \emph{J. Chem. Phys.} \textbf{2016}, \emph{144}, 234102\relax
\mciteBstWouldAddEndPuncttrue
\mciteSetBstMidEndSepPunct{\mcitedefaultmidpunct}
{\mcitedefaultendpunct}{\mcitedefaultseppunct}\relax
\EndOfBibitem
\bibitem[Goings \latin{et~al.}(2016)Goings, Kasper, Egidi, Sun, and
  Li]{Li16_104107}
Goings,~J.~J.; Kasper,~J.~M.; Egidi,~F.; Sun,~S.; Li,~X. Real Time Propagation
  of the Exact Two Component Time-Dependent Density Functional Theory. \emph{J.
  Chem. Phys.} \textbf{2016}, \emph{145}, 104107\relax
\mciteBstWouldAddEndPuncttrue
\mciteSetBstMidEndSepPunct{\mcitedefaultmidpunct}
{\mcitedefaultendpunct}{\mcitedefaultseppunct}\relax
\EndOfBibitem
\bibitem[Kasper \latin{et~al.}(2018)Kasper, Lestrange, Stetina, and
  Li]{Li18_1998}
Kasper,~J.~M.; Lestrange,~P.~J.; Stetina,~T.~F.; Li,~X. Modeling L$_{2,3}$-Edge
  X-ray Absorption Spectroscopy with Real-Time Exact Two-Component Relativistic
  Time-Dependent Density Functional Theory. \emph{J. Chem. Theory Comput.}
  \textbf{2018}, \emph{14}, 1998--2006\relax
\mciteBstWouldAddEndPuncttrue
\mciteSetBstMidEndSepPunct{\mcitedefaultmidpunct}
{\mcitedefaultendpunct}{\mcitedefaultseppunct}\relax
\EndOfBibitem
\bibitem[Stetina \latin{et~al.}(2019)Stetina, Kasper, and Li]{Li19_234103}
Stetina,~T.~F.; Kasper,~J.~M.; Li,~X. Modeling L$_2,_3$-Edge X-ray Absorption
  Spectroscopy with Linear Response Exact Two-Component Relativistic
  Time-Dependent Density Functional Theory. \emph{J. Chem. Phys.}
  \textbf{2019}, \emph{150}, 234103\relax
\mciteBstWouldAddEndPuncttrue
\mciteSetBstMidEndSepPunct{\mcitedefaultmidpunct}
{\mcitedefaultendpunct}{\mcitedefaultseppunct}\relax
\EndOfBibitem
\bibitem[Sonk \latin{et~al.}(2011)Sonk, Caricato, and
  Schlegel]{Schlegel11_4678}
Sonk,~J.~A.; Caricato,~M.; Schlegel,~H.~B. {TD-CI Simulation of the Electronic
  Optical Response of Molecules in Intense Fields: Comparison of RPA, CIS,
  CIS(D), and EOM-CCSD}. \emph{J. Phys. Chem. A} \textbf{2011}, \emph{115},
  4678--4690\relax
\mciteBstWouldAddEndPuncttrue
\mciteSetBstMidEndSepPunct{\mcitedefaultmidpunct}
{\mcitedefaultendpunct}{\mcitedefaultseppunct}\relax
\EndOfBibitem
\bibitem[Luppi and Head-Gordon(2012)Luppi, and Head-Gordon]{HeadGordon12_909}
Luppi,~E.; Head-Gordon,~M. {Computation of High-harmonic Generation Spectra of
  H$_2$ and N$_2$ in Intense Laser Pulses Using Quantum Chemistry Methods and
  Time-dependent Density Functional Theory}. \emph{Mol. Phys.} \textbf{2012},
  \emph{110}, 909--923\relax
\mciteBstWouldAddEndPuncttrue
\mciteSetBstMidEndSepPunct{\mcitedefaultmidpunct}
{\mcitedefaultendpunct}{\mcitedefaultseppunct}\relax
\EndOfBibitem
\bibitem[Pawłowski \latin{et~al.}(2015)Pawłowski, Olsen, and
  J{\o}rgensen]{Jorgensen15_114109}
Pawłowski,~F.; Olsen,~J.; J{\o}rgensen,~P. Molecular response properties from
  a Hermitian eigenvalue equation for a time-periodic Hamiltonian. \emph{J.
  Chem. Phys.} \textbf{2015}, \emph{142}, 114109\relax
\mciteBstWouldAddEndPuncttrue
\mciteSetBstMidEndSepPunct{\mcitedefaultmidpunct}
{\mcitedefaultendpunct}{\mcitedefaultseppunct}\relax
\EndOfBibitem
\bibitem[Coriani \latin{et~al.}(2016)Coriani, Pawłowski, Olsen, and
  J{\o}rgensen]{Jorgensen16_024102}
Coriani,~S.; Pawłowski,~F.; Olsen,~J.; J{\o}rgensen,~P. Molecular response
  properties in equation of motion coupled cluster theory: A time-dependent
  perspective. \emph{J. Chem. Phys.} \textbf{2016}, \emph{144}, 024102\relax
\mciteBstWouldAddEndPuncttrue
\mciteSetBstMidEndSepPunct{\mcitedefaultmidpunct}
{\mcitedefaultendpunct}{\mcitedefaultseppunct}\relax
\EndOfBibitem
\bibitem[Nascimento and DePrince(2016)Nascimento, and
  DePrince]{DePrince16_5834}
Nascimento,~D.~R.; DePrince,~A.~E. Linear Absorption Spectra from Explicitly
  Time-Dependent Equation-of-Motion Coupled-Cluster Theory. \emph{J. Chem.
  Theory Comput.} \textbf{2016}, \emph{12}, 5834--5840\relax
\mciteBstWouldAddEndPuncttrue
\mciteSetBstMidEndSepPunct{\mcitedefaultmidpunct}
{\mcitedefaultendpunct}{\mcitedefaultseppunct}\relax
\EndOfBibitem
\bibitem[Nascimento and DePrince(2017)Nascimento, and
  DePrince]{DePrince17_2951}
Nascimento,~D.~R.; DePrince,~A.~E. Simulation of Near-Edge X-ray Absorption
  Fine Structure with Time-Dependent Equation-of-Motion Coupled-Cluster Theory.
  \emph{J. Phys. Chem. Lett.} \textbf{2017}, \emph{8}, 2951--2957\relax
\mciteBstWouldAddEndPuncttrue
\mciteSetBstMidEndSepPunct{\mcitedefaultmidpunct}
{\mcitedefaultendpunct}{\mcitedefaultseppunct}\relax
\EndOfBibitem
\bibitem[Koulias \latin{et~al.}(2019)Koulias, Williams-Young, Nascimento,
  DePrince, and Li]{Li19_6617}
Koulias,~L.~N.; Williams-Young,~D.~B.; Nascimento,~D.~R.; DePrince,~A.~E.;
  Li,~X. Relativistic Time-Dependent Equation-of-Motion Coupled-Cluster.
  \emph{J. Chem. Theory Comput.} \textbf{2019}, \emph{15}, 6617--6624\relax
\mciteBstWouldAddEndPuncttrue
\mciteSetBstMidEndSepPunct{\mcitedefaultmidpunct}
{\mcitedefaultendpunct}{\mcitedefaultseppunct}\relax
\EndOfBibitem
\bibitem[Cooper \latin{et~al.}(2021)Cooper, Koulias, Nascimento, Li, and
  DePrince]{Li21_5438}
Cooper,~B.~C.; Koulias,~L.~N.; Nascimento,~D.~R.; Li,~X.; DePrince,~A.~E. Short
  Iterative Lanczos Integration in Time-Dependent Equation-of-Motion
  Coupled-Cluster Theory. \emph{J. Phys. Chem. A} \textbf{2021}, \emph{125},
  5438--5447\relax
\mciteBstWouldAddEndPuncttrue
\mciteSetBstMidEndSepPunct{\mcitedefaultmidpunct}
{\mcitedefaultendpunct}{\mcitedefaultseppunct}\relax
\EndOfBibitem
\bibitem[Yuwono \latin{et~al.}(2023)Yuwono, Cooper, Zhang, Li, and
  DePrince]{Li23_044113}
Yuwono,~S.~H.; Cooper,~B.~C.; Zhang,~T.; Li,~X.; DePrince,~I.,~A.~Eugene
  {Time-Dependent Equation-of-Motion Coupled-Cluster Simulations with a
  Defective Hamiltonian}. \emph{J. Chem. Phys.} \textbf{2023}, \emph{159},
  044113\relax
\mciteBstWouldAddEndPuncttrue
\mciteSetBstMidEndSepPunct{\mcitedefaultmidpunct}
{\mcitedefaultendpunct}{\mcitedefaultseppunct}\relax
\EndOfBibitem
\bibitem[Fales and Levine(2015)Fales, and Levine]{Levine15_4708}
Fales,~B.~S.; Levine,~B.~G. {Nanoscale Multireference Quantum Chemistry: Full
  Configuration Interaction on Graphical Processing Units}. \emph{J. Chem.
  Theory Comput.} \textbf{2015}, \emph{11}, 4708--4716\relax
\mciteBstWouldAddEndPuncttrue
\mciteSetBstMidEndSepPunct{\mcitedefaultmidpunct}
{\mcitedefaultendpunct}{\mcitedefaultseppunct}\relax
\EndOfBibitem
\bibitem[Shu \latin{et~al.}(2015)Shu, Hohenstein, and Levine]{Levine15_024102}
Shu,~Y.; Hohenstein,~E.~G.; Levine,~B.~G. {Configuration Interaction Singles
  Natural Orbitals: An Orbital Basis for an Efficient and size Intensive
  Multireference Description of Electronic Excited States}. \emph{J. Chem.
  Phys.} \textbf{2015}, \emph{142}, 024102\relax
\mciteBstWouldAddEndPuncttrue
\mciteSetBstMidEndSepPunct{\mcitedefaultmidpunct}
{\mcitedefaultendpunct}{\mcitedefaultseppunct}\relax
\EndOfBibitem
\bibitem[Lestrange \latin{et~al.}(2018)Lestrange, Hoffmann, and Li]{Li18_295}
Lestrange,~P.~J.; Hoffmann,~M.~R.; Li,~X. Time-Dependent Configuration
  Interaction using the Graphical Unitary Group Approach: Nonlinear Electric
  Properties. \emph{Adv. Quantum Chem.} \textbf{2018}, \emph{76},
  295--313\relax
\mciteBstWouldAddEndPuncttrue
\mciteSetBstMidEndSepPunct{\mcitedefaultmidpunct}
{\mcitedefaultendpunct}{\mcitedefaultseppunct}\relax
\EndOfBibitem
\bibitem[Peng \latin{et~al.}(2018)Peng, Fales, and Levine]{Levine18_4129}
Peng,~W.~T.; Fales,~B.~S.; Levine,~B.~G. {Simulating Electron Dynamics of
  Complex Molecules with Time-Dependent Complete Active Space Configuration
  Interaction}. \emph{J. Chem. Theory Comput.} \textbf{2018}, \emph{14},
  4129--4138\relax
\mciteBstWouldAddEndPuncttrue
\mciteSetBstMidEndSepPunct{\mcitedefaultmidpunct}
{\mcitedefaultendpunct}{\mcitedefaultseppunct}\relax
\EndOfBibitem
\bibitem[Ulusoy \latin{et~al.}(2018)Ulusoy, Stewart, and
  Wilson]{Wilson18_014107}
Ulusoy,~I.~S.; Stewart,~Z.; Wilson,~A.~K. {The Role of the CI Expansion Length
  in Time-Dependent Studies}. \emph{J. Chem. Phys.} \textbf{2018}, \emph{148},
  014107\relax
\mciteBstWouldAddEndPuncttrue
\mciteSetBstMidEndSepPunct{\mcitedefaultmidpunct}
{\mcitedefaultendpunct}{\mcitedefaultseppunct}\relax
\EndOfBibitem
\bibitem[Liu \latin{et~al.}(2019)Liu, Jenkins, Wildman, Frisch, Lipparini,
  Mennucci, and Li]{Li19_1633}
Liu,~H.; Jenkins,~A.~J.; Wildman,~A.; Frisch,~M.~J.; Lipparini,~F.;
  Mennucci,~B.; Li,~X. Time-Dependent Complete Active Space Embedded in a
  Polarizable Force Field. \emph{J. Chem. Theory Comput.} \textbf{2019},
  \emph{15}, 1633--1641\relax
\mciteBstWouldAddEndPuncttrue
\mciteSetBstMidEndSepPunct{\mcitedefaultmidpunct}
{\mcitedefaultendpunct}{\mcitedefaultseppunct}\relax
\EndOfBibitem
\bibitem[Goings \latin{et~al.}(2018)Goings, Lestrange, and Li]{Li18_e1341}
Goings,~J.~J.; Lestrange,~P.~J.; Li,~X. Real-Time Time-Dependent Electronic
  Structure Theory. \emph{WIREs Comput. Mol. Sci.} \textbf{2018}, \emph{8},
  e1341\relax
\mciteBstWouldAddEndPuncttrue
\mciteSetBstMidEndSepPunct{\mcitedefaultmidpunct}
{\mcitedefaultendpunct}{\mcitedefaultseppunct}\relax
\EndOfBibitem
\bibitem[Li \latin{et~al.}(2020)Li, Govind, Isborn, DePrince, and
  Lopata]{Li20_9951}
Li,~X.; Govind,~N.; Isborn,~C.; DePrince,~A.~E.; Lopata,~K. Real-Time
  Time-Dependent Electronic Structure Theory. \emph{Chem. Rev.} \textbf{2020},
  \emph{120}, 9951--9993\relax
\mciteBstWouldAddEndPuncttrue
\mciteSetBstMidEndSepPunct{\mcitedefaultmidpunct}
{\mcitedefaultendpunct}{\mcitedefaultseppunct}\relax
\EndOfBibitem
\bibitem[Rai \latin{et~al.}(2011)Rai, Hod, and Nitzan]{Nitzan11_2118}
Rai,~D.; Hod,~O.; Nitzan,~A. Magnetic Field Control of the Current through
  Molecular Ring Junctions. \emph{The Journal of Physical Chemistry Letters}
  \textbf{2011}, \emph{2}, 2118--2124\relax
\mciteBstWouldAddEndPuncttrue
\mciteSetBstMidEndSepPunct{\mcitedefaultmidpunct}
{\mcitedefaultendpunct}{\mcitedefaultseppunct}\relax
\EndOfBibitem
\bibitem[Rai \latin{et~al.}(2012)Rai, Hod, and Nitzan]{Nitzan12_155440}
Rai,~D.; Hod,~O.; Nitzan,~A. Magnetic fields effects on the electronic
  conduction properties of molecular ring structures. \emph{Phys. Rev. B}
  \textbf{2012}, \emph{85}, 155440\relax
\mciteBstWouldAddEndPuncttrue
\mciteSetBstMidEndSepPunct{\mcitedefaultmidpunct}
{\mcitedefaultendpunct}{\mcitedefaultseppunct}\relax
\EndOfBibitem
\bibitem[Sundholm \latin{et~al.}(2021)Sundholm, Dimitrova, and
  Berger]{Berger21_12362}
Sundholm,~D.; Dimitrova,~M.; Berger,~R. J.~F. Current Density and Molecular
  Magnetic Properties. \emph{Chem. Commun.} \textbf{2021}, \emph{57},
  12362--12378\relax
\mciteBstWouldAddEndPuncttrue
\mciteSetBstMidEndSepPunct{\mcitedefaultmidpunct}
{\mcitedefaultendpunct}{\mcitedefaultseppunct}\relax
\EndOfBibitem
\bibitem[Bro-Jørgensen \latin{et~al.}(2025)Bro-Jørgensen, Sauer, Solomon, and
  Garner]{Garner25_4073}
Bro-Jørgensen,~W.; Sauer,~S. P.~A.; Solomon,~G.~C.; Garner,~M.~H. Substantial
  Magnetic Fields Arising from Ballistic Ring Currents in Single-Molecule
  Junctions. \emph{JACS Au} \textbf{2025}, \emph{5}, 4073--4085\relax
\mciteBstWouldAddEndPuncttrue
\mciteSetBstMidEndSepPunct{\mcitedefaultmidpunct}
{\mcitedefaultendpunct}{\mcitedefaultseppunct}\relax
\EndOfBibitem
\bibitem[Maaravi and Hod(2020)Maaravi, and Hod]{Hod20_8652}
Maaravi,~T.; Hod,~O. Simulating Electron Dynamics in Open Quantum Systems under
  Magnetic Fields. \emph{The Journal of Physical Chemistry C} \textbf{2020},
  \emph{124}, 8652--8662\relax
\mciteBstWouldAddEndPuncttrue
\mciteSetBstMidEndSepPunct{\mcitedefaultmidpunct}
{\mcitedefaultendpunct}{\mcitedefaultseppunct}\relax
\EndOfBibitem
\bibitem[Jusélius \latin{et~al.}(2004)Jusélius, Sundholm, and
  Gauss]{Gauss04_3952}
Jusélius,~J.; Sundholm,~D.; Gauss,~J. Calculation of current densities using
  gauge-including atomic orbitals. \emph{J. Chem. Phys.} \textbf{2004},
  \emph{121}, 3952--3963\relax
\mciteBstWouldAddEndPuncttrue
\mciteSetBstMidEndSepPunct{\mcitedefaultmidpunct}
{\mcitedefaultendpunct}{\mcitedefaultseppunct}\relax
\EndOfBibitem
\bibitem[Fliegl \latin{et~al.}(2011)Fliegl, Taubert, Lehtonen, and
  Sundholm]{Sundholm11_20500}
Fliegl,~H.; Taubert,~S.; Lehtonen,~O.; Sundholm,~D. The Gauge Including
  Magnetically Induced Current Method. \emph{Phys. Chem. Chem. Phys.}
  \textbf{2011}, \emph{13}, 20500--20518\relax
\mciteBstWouldAddEndPuncttrue
\mciteSetBstMidEndSepPunct{\mcitedefaultmidpunct}
{\mcitedefaultendpunct}{\mcitedefaultseppunct}\relax
\EndOfBibitem
\bibitem[Sun \latin{et~al.}(2021)Sun, Stetina, Zhang, Hu, Valeev, Sun, and
  Li]{Li21_3388}
Sun,~S.; Stetina,~T.~F.; Zhang,~T.; Hu,~H.; Valeev,~E.~F.; Sun,~Q.; Li,~X.
  Efficient Four-Component Dirac--Coulomb--Gaunt Hartree--Fock in the Pauli
  Spinor Representation. \emph{J. Chem. Theory Comput.} \textbf{2021},
  \emph{17}, 3388--3402\relax
\mciteBstWouldAddEndPuncttrue
\mciteSetBstMidEndSepPunct{\mcitedefaultmidpunct}
{\mcitedefaultendpunct}{\mcitedefaultseppunct}\relax
\EndOfBibitem
\bibitem[Sun \latin{et~al.}(2022)Sun, Ehrman, Sun, and Li]{Li22_064112}
Sun,~S.; Ehrman,~J.~N.; Sun,~Q.; Li,~X. Efficient Evaluation of the Breit
  Operator in the Pauli Spinor Basis. \emph{J. Chem. Phys.} \textbf{2022},
  \emph{157}, 064112\relax
\mciteBstWouldAddEndPuncttrue
\mciteSetBstMidEndSepPunct{\mcitedefaultmidpunct}
{\mcitedefaultendpunct}{\mcitedefaultseppunct}\relax
\EndOfBibitem
\bibitem[Sun \latin{et~al.}(2023)Sun, Ehrman, Zhang, Sun, Dyall, and
  Li]{Li23_171101}
Sun,~S.; Ehrman,~J.; Zhang,~T.; Sun,~Q.; Dyall,~K.~G.; Li,~X. {Scalar Breit
  Interaction for Molecular Calculations}. \emph{J. Chem. Phys.} \textbf{2023},
  \emph{158}, 171101\relax
\mciteBstWouldAddEndPuncttrue
\mciteSetBstMidEndSepPunct{\mcitedefaultmidpunct}
{\mcitedefaultendpunct}{\mcitedefaultseppunct}\relax
\EndOfBibitem
\bibitem[Bast \latin{et~al.}(2009)Bast, Jusélius, and Saue]{Saue09_187}
Bast,~R.; Jusélius,~J.; Saue,~T. 4-Component Relativistic Calculation of the
  Magnetically Induced Current Density in the Group 15 Heteroaromatic
  Compounds. \emph{Chem. Phys.} \textbf{2009}, \emph{356}, 187--194\relax
\mciteBstWouldAddEndPuncttrue
\mciteSetBstMidEndSepPunct{\mcitedefaultmidpunct}
{\mcitedefaultendpunct}{\mcitedefaultseppunct}\relax
\EndOfBibitem
\bibitem[Sulzer \latin{et~al.}(2011)Sulzer, Olejniczak, Bast, and
  Saue]{Saue11_20682}
Sulzer,~D.; Olejniczak,~M.; Bast,~R.; Saue,~T. 4-Component Relativistic
  Magnetically Induced Current Density using London Atomic Orbitals.
  \emph{Phys. Chem. Chem. Phys.} \textbf{2011}, \emph{13}, 20682--20689\relax
\mciteBstWouldAddEndPuncttrue
\mciteSetBstMidEndSepPunct{\mcitedefaultmidpunct}
{\mcitedefaultendpunct}{\mcitedefaultseppunct}\relax
\EndOfBibitem
\bibitem[Schaffhauser and K\"ummel(2016)Schaffhauser, and
  K\"ummel]{Kummel16_035115}
Schaffhauser,~P.; K\"ummel,~S. Using time-dependent density functional theory
  in real time for calculating electronic transport. \emph{Phys. Rev. B}
  \textbf{2016}, \emph{93}, 035115\relax
\mciteBstWouldAddEndPuncttrue
\mciteSetBstMidEndSepPunct{\mcitedefaultmidpunct}
{\mcitedefaultendpunct}{\mcitedefaultseppunct}\relax
\EndOfBibitem
\bibitem[Bushong \latin{et~al.}(2005)Bushong, Sai, and
  Di~Ventra]{DiVentra05_2569}
Bushong,~N.; Sai,~N.; Di~Ventra,~M. Approach to Steady-State Transport in
  Nanoscale Conductors. \emph{Nano Letters} \textbf{2005}, \emph{5},
  2569--2572\relax
\mciteBstWouldAddEndPuncttrue
\mciteSetBstMidEndSepPunct{\mcitedefaultmidpunct}
{\mcitedefaultendpunct}{\mcitedefaultseppunct}\relax
\EndOfBibitem
\bibitem[Cheng \latin{et~al.}(2006)Cheng, Evans, and
  Van~Voorhis]{VanVoorhis06_155112}
Cheng,~C.-L.; Evans,~J.~S.; Van~Voorhis,~T. Simulating molecular conductance
  using real-time density functional theory. \emph{Phys. Rev. B} \textbf{2006},
  \emph{74}, 155112\relax
\mciteBstWouldAddEndPuncttrue
\mciteSetBstMidEndSepPunct{\mcitedefaultmidpunct}
{\mcitedefaultendpunct}{\mcitedefaultseppunct}\relax
\EndOfBibitem
\bibitem[Dyall(1997)]{Dyall97_9618}
Dyall,~K.~G. Interfacing Relativistic and Nonrelativistic Methods. I.
  Normalized Elimination of the Small Component in the Modified Dirac Equation.
  \emph{J. Chem. Phys.} \textbf{1997}, \emph{106}, 9618--9626\relax
\mciteBstWouldAddEndPuncttrue
\mciteSetBstMidEndSepPunct{\mcitedefaultmidpunct}
{\mcitedefaultendpunct}{\mcitedefaultseppunct}\relax
\EndOfBibitem
\bibitem[Dyall(1998)]{Dyall98_4201}
Dyall,~K.~G. Interfacing Relativistic and Nonrelativistic Methods. {II.}
  Investigation of a Low-Order Approximation. \emph{J. Chem. Phys.}
  \textbf{1998}, \emph{109}, 4201--4208\relax
\mciteBstWouldAddEndPuncttrue
\mciteSetBstMidEndSepPunct{\mcitedefaultmidpunct}
{\mcitedefaultendpunct}{\mcitedefaultseppunct}\relax
\EndOfBibitem
\bibitem[Dyall and Enevoldsen(1999)Dyall, and Enevoldsen]{Enevoldsen99_10000}
Dyall,~K.~G.; Enevoldsen,~T. Interfacing Relativistic and Nonrelativistic
  Methods. {III.} Atomic 4-Spinor Expansions and Integral Approximations.
  \emph{J. Chem. Phys.} \textbf{1999}, \emph{111}, 10000--10007\relax
\mciteBstWouldAddEndPuncttrue
\mciteSetBstMidEndSepPunct{\mcitedefaultmidpunct}
{\mcitedefaultendpunct}{\mcitedefaultseppunct}\relax
\EndOfBibitem
\bibitem[Dyall(2001)]{Dyall01_9136}
Dyall,~K.~G. Interfacing Relativistic and Nonrelativistic Methods. {IV..}
  {O}ne- and Two-Electron Scalar Approximations. \emph{J. Chem. Phys.}
  \textbf{2001}, \emph{115}, 9136--9143\relax
\mciteBstWouldAddEndPuncttrue
\mciteSetBstMidEndSepPunct{\mcitedefaultmidpunct}
{\mcitedefaultendpunct}{\mcitedefaultseppunct}\relax
\EndOfBibitem
\bibitem[Kutzlenigg and Liu(2005)Kutzlenigg, and Liu]{Liu05_241102}
Kutzlenigg,~W.; Liu,~W. Quasirelativistic Theory Equivalent to Fully
  Relativistic Theory. \emph{J. Chem. Phys.} \textbf{2005}, \emph{123},
  241102\relax
\mciteBstWouldAddEndPuncttrue
\mciteSetBstMidEndSepPunct{\mcitedefaultmidpunct}
{\mcitedefaultendpunct}{\mcitedefaultseppunct}\relax
\EndOfBibitem
\bibitem[Liu and Peng(2006)Liu, and Peng]{Peng06_044102}
Liu,~W.; Peng,~D. Infinite-Order Quasirelativistic Density Functional Method
  Based on the Exact Matrix Quasirelativistic Theory. \emph{J. Chem. Phys.}
  \textbf{2006}, \emph{125}, 044102\relax
\mciteBstWouldAddEndPuncttrue
\mciteSetBstMidEndSepPunct{\mcitedefaultmidpunct}
{\mcitedefaultendpunct}{\mcitedefaultseppunct}\relax
\EndOfBibitem
\bibitem[Peng \latin{et~al.}(2007)Peng, Liu, Xiao, and Cheng]{Cheng07_104106}
Peng,~D.; Liu,~W.; Xiao,~Y.; Cheng,~L. Making Four- and Two-Component
  Relativistic Density Functional Methods Fully Equivalent Based on the Idea of
  From Atoms to Molecule. \emph{J. Chem. Phys.} \textbf{2007}, \emph{127},
  104106\relax
\mciteBstWouldAddEndPuncttrue
\mciteSetBstMidEndSepPunct{\mcitedefaultmidpunct}
{\mcitedefaultendpunct}{\mcitedefaultseppunct}\relax
\EndOfBibitem
\bibitem[Ilias and Saue(2007)Ilias, and Saue]{Saue07_064102}
Ilias,~M.; Saue,~T. An Infinite-Order Relativistic Hamiltonian by a Simple
  One-Step Transformation. \emph{J. Chem. Phys.} \textbf{2007}, \emph{126},
  064102\relax
\mciteBstWouldAddEndPuncttrue
\mciteSetBstMidEndSepPunct{\mcitedefaultmidpunct}
{\mcitedefaultendpunct}{\mcitedefaultseppunct}\relax
\EndOfBibitem
\bibitem[Liu and Peng(2009)Liu, and Peng]{Peng09_031104}
Liu,~W.; Peng,~D. Exact Two-component Hamiltonians Revisited. \emph{J. Chem.
  Phys.} \textbf{2009}, \emph{131}, 031104\relax
\mciteBstWouldAddEndPuncttrue
\mciteSetBstMidEndSepPunct{\mcitedefaultmidpunct}
{\mcitedefaultendpunct}{\mcitedefaultseppunct}\relax
\EndOfBibitem
\bibitem[Liu(2010)]{Liu10_1679}
Liu,~W. Ideas of Relativistic Quantum Chemistry. \emph{Mol. Phys.}
  \textbf{2010}, \emph{108}, 1679--1706\relax
\mciteBstWouldAddEndPuncttrue
\mciteSetBstMidEndSepPunct{\mcitedefaultmidpunct}
{\mcitedefaultendpunct}{\mcitedefaultseppunct}\relax
\EndOfBibitem
\bibitem[Li \latin{et~al.}(2012)Li, Xiao, and Liu]{Liu12_154114}
Li,~Z.; Xiao,~Y.; Liu,~W. On the Spin Separation of Algebraic Two-Component
  Relativistic Hamiltonians. \emph{J. Chem. Phys.} \textbf{2012}, \emph{137},
  154114\relax
\mciteBstWouldAddEndPuncttrue
\mciteSetBstMidEndSepPunct{\mcitedefaultmidpunct}
{\mcitedefaultendpunct}{\mcitedefaultseppunct}\relax
\EndOfBibitem
\bibitem[Peng \latin{et~al.}(2013)Peng, Middendorf, Weigend, and
  Reiher]{Reiher13_184105}
Peng,~D.; Middendorf,~N.; Weigend,~F.; Reiher,~M. An Efficient Implementation
  of Two-Component Relativistic Exact-Decoupling Methods for Large Molecules.
  \emph{J. Chem. Phys.} \textbf{2013}, \emph{138}, 184105\relax
\mciteBstWouldAddEndPuncttrue
\mciteSetBstMidEndSepPunct{\mcitedefaultmidpunct}
{\mcitedefaultendpunct}{\mcitedefaultseppunct}\relax
\EndOfBibitem
\bibitem[Egidi \latin{et~al.}(2016)Egidi, Goings, Frisch, and Li]{Li16_3711}
Egidi,~F.; Goings,~J.~J.; Frisch,~M.~J.; Li,~X. Direct Atomic-Orbital-Based
  Relativistic Two-Component Linear Response Method for Calculating
  Excited-State Fine Structures. \emph{J. Chem. Theory Comput.} \textbf{2016},
  \emph{12}, 3711--3718\relax
\mciteBstWouldAddEndPuncttrue
\mciteSetBstMidEndSepPunct{\mcitedefaultmidpunct}
{\mcitedefaultendpunct}{\mcitedefaultseppunct}\relax
\EndOfBibitem
\bibitem[Konecny \latin{et~al.}(2016)Konecny, Kadek, Komorovsky, Malkina, Ruud,
  and Repisky]{Repisky16_5823}
Konecny,~L.; Kadek,~M.; Komorovsky,~S.; Malkina,~O.~L.; Ruud,~K.; Repisky,~M.
  Acceleration of Relativistic Electron Dynamics by Means of {X2C}
  Transformation: {A}pplication to the Calculation of Nonlinear Optical
  Properties. \emph{J. Chem. Theory Comput.} \textbf{2016}, \emph{12},
  5823--5833\relax
\mciteBstWouldAddEndPuncttrue
\mciteSetBstMidEndSepPunct{\mcitedefaultmidpunct}
{\mcitedefaultendpunct}{\mcitedefaultseppunct}\relax
\EndOfBibitem
\bibitem[Egidi \latin{et~al.}(2017)Egidi, Sun, Goings, Scalmani, Frisch, and
  Li]{Li17_2591}
Egidi,~F.; Sun,~S.; Goings,~J.~J.; Scalmani,~G.; Frisch,~M.~J.; Li,~X.
  Two-Component Non-Collinear Time-Dependent Spin Density Functional Theory for
  Excited State Calculations. \emph{J. Chem. Theory Comput.} \textbf{2017},
  \emph{13}, 2591--2603\relax
\mciteBstWouldAddEndPuncttrue
\mciteSetBstMidEndSepPunct{\mcitedefaultmidpunct}
{\mcitedefaultendpunct}{\mcitedefaultseppunct}\relax
\EndOfBibitem
\bibitem[Liu and Cheng(2021)Liu, and Cheng]{Cheng21_1536}
Liu,~J.; Cheng,~L. Relativistic Coupled-Cluster and Equation-of-Motion
  Coupled-Cluster Methods. \emph{WIREs Comput. Mol. Sci.} \textbf{2021},
  \emph{11}, 1536\relax
\mciteBstWouldAddEndPuncttrue
\mciteSetBstMidEndSepPunct{\mcitedefaultmidpunct}
{\mcitedefaultendpunct}{\mcitedefaultseppunct}\relax
\EndOfBibitem
\bibitem[Sharma \latin{et~al.}(2022)Sharma, Jenkins, Scalmani, Frisch, Truhlar,
  Gagliardi, and Li]{Li22_2947}
Sharma,~P.; Jenkins,~A.~J.; Scalmani,~G.; Frisch,~M.~J.; Truhlar,~D.~G.;
  Gagliardi,~L.; Li,~X. Exact-Two-Component Multiconfiguration Pair-Density
  Functional Theory. \emph{J. Chem. Theory Comput.} \textbf{2022}, \emph{18},
  2947--2954\relax
\mciteBstWouldAddEndPuncttrue
\mciteSetBstMidEndSepPunct{\mcitedefaultmidpunct}
{\mcitedefaultendpunct}{\mcitedefaultseppunct}\relax
\EndOfBibitem
\bibitem[Lu \latin{et~al.}(2022)Lu, Hu, Jenkins, and Li]{Li22_2983}
Lu,~L.; Hu,~H.; Jenkins,~A.~J.; Li,~X. Exact-Two-Component Relativistic
  Multireference Second-Order Perturbation Theory. \emph{J. Chem. Theory
  Comput.} \textbf{2022}, \emph{18}, 2983--2992\relax
\mciteBstWouldAddEndPuncttrue
\mciteSetBstMidEndSepPunct{\mcitedefaultmidpunct}
{\mcitedefaultendpunct}{\mcitedefaultseppunct}\relax
\EndOfBibitem
\bibitem[Hoyer \latin{et~al.}(2022)Hoyer, Hu, Lu, Knecht, and Li]{Li22_5011}
Hoyer,~C.~E.; Hu,~H.; Lu,~L.; Knecht,~S.; Li,~X. Relativistic
  Kramers-Unrestricted Exact-Two-Component Density Matrix Renormalization
  Group. \emph{J. Phys. Chem. A} \textbf{2022}, \emph{126}, 5011--5020\relax
\mciteBstWouldAddEndPuncttrue
\mciteSetBstMidEndSepPunct{\mcitedefaultmidpunct}
{\mcitedefaultendpunct}{\mcitedefaultseppunct}\relax
\EndOfBibitem
\bibitem[Knecht \latin{et~al.}(2022)Knecht, Repisky, Jensen, and
  Saue]{Saue22_114106}
Knecht,~S.; Repisky,~M.; Jensen,~H. J.~{\relax Aa}.; Saue,~T. {Exact
  Two-component Hamiltonians for Relativistic Quantum Chemistry: Two-electron
  Picture-change Corrections Made Simple}. \emph{J. Chem. Phys.} \textbf{2022},
  \emph{157}, 114106\relax
\mciteBstWouldAddEndPuncttrue
\mciteSetBstMidEndSepPunct{\mcitedefaultmidpunct}
{\mcitedefaultendpunct}{\mcitedefaultseppunct}\relax
\EndOfBibitem
\bibitem[Ehrman \latin{et~al.}(2023)Ehrman, Martinez-Baez, Jenkins, and
  Li]{Li23_5785}
Ehrman,~J.; Martinez-Baez,~E.; Jenkins,~A.~J.; Li,~X. Improving One-Electron
  Exact-Two-Component Relativistic Methods with the
  Dirac--Coulomb--Breit-Parameterized Effective Spin--Orbit Coupling. \emph{J.
  Chem. Theory Comput.} \textbf{2023}, \emph{19}, 5785--5790\relax
\mciteBstWouldAddEndPuncttrue
\mciteSetBstMidEndSepPunct{\mcitedefaultmidpunct}
{\mcitedefaultendpunct}{\mcitedefaultseppunct}\relax
\EndOfBibitem
\bibitem[Capelle \latin{et~al.}(2001)Capelle, Vignale, and
  Gy{\"{o}}rffy]{Gyorffy01_206403}
Capelle,~K.; Vignale,~G.; Gy{\"{o}}rffy,~B. Spin Currents and Spin Dynamics in
  Time-Dependent Densit-yFunctional Theory. \emph{Phys. Rev. Lett.}
  \textbf{2001}, \emph{87}, 206403\relax
\mciteBstWouldAddEndPuncttrue
\mciteSetBstMidEndSepPunct{\mcitedefaultmidpunct}
{\mcitedefaultendpunct}{\mcitedefaultseppunct}\relax
\EndOfBibitem
\bibitem[van W\"{u}llen(2002)]{vanWullen02_779}
van W\"{u}llen,~C. Spin Densities in Two-Component Relativistic Density
  Functional Calculations: Noncollinear versus Collinear Approach. \emph{J.
  Comput. Chem.} \textbf{2002}, \emph{23}, 779--785\relax
\mciteBstWouldAddEndPuncttrue
\mciteSetBstMidEndSepPunct{\mcitedefaultmidpunct}
{\mcitedefaultendpunct}{\mcitedefaultseppunct}\relax
\EndOfBibitem
\bibitem[Wang and Liu(2003)Wang, and Liu]{Liu03_597}
Wang,~F.; Liu,~W. Comparison of Different Polarization Schemes in Open-shell
  Relativistic Density Functional Calculations. \emph{J. Chin. Chem. Soc.}
  \textbf{2003}, \emph{50}, 597--606\relax
\mciteBstWouldAddEndPuncttrue
\mciteSetBstMidEndSepPunct{\mcitedefaultmidpunct}
{\mcitedefaultendpunct}{\mcitedefaultseppunct}\relax
\EndOfBibitem
\bibitem[Anton \latin{et~al.}(2004)Anton, Fricke, and Engel]{Engel04_012505}
Anton,~J.; Fricke,~B.; Engel,~E. Noncollinear and Collinear Relativistic
  Density-functional Program for Electric and Magnetic Properties of Molecules.
  \emph{Phys. Rev. A} \textbf{2004}, \emph{69}, 012505\relax
\mciteBstWouldAddEndPuncttrue
\mciteSetBstMidEndSepPunct{\mcitedefaultmidpunct}
{\mcitedefaultendpunct}{\mcitedefaultseppunct}\relax
\EndOfBibitem
\bibitem[Wang and Ziegler(2004)Wang, and Ziegler]{Ziegler04_12191}
Wang,~F.; Ziegler,~T. Time-Dependent Density Functional Theory Based on a
  Noncollinear Formulation of the Exchange-Correlation Potential. \emph{J.
  Chem. Phys.} \textbf{2004}, \emph{121}, 12191--12196\relax
\mciteBstWouldAddEndPuncttrue
\mciteSetBstMidEndSepPunct{\mcitedefaultmidpunct}
{\mcitedefaultendpunct}{\mcitedefaultseppunct}\relax
\EndOfBibitem
\bibitem[Gao \latin{et~al.}(2004)Gao, , Liu, Song, and Liu]{Liu04_6658}
Gao,~J.; ; Liu,~W.; Song,~B.; Liu,~C. Time-Dependent Four-Component
  Relativistic Density Functional Theory for Excitation Energies. \emph{J.
  Chem. Phys.} \textbf{2004}, \emph{121}, 6658--6666\relax
\mciteBstWouldAddEndPuncttrue
\mciteSetBstMidEndSepPunct{\mcitedefaultmidpunct}
{\mcitedefaultendpunct}{\mcitedefaultseppunct}\relax
\EndOfBibitem
\bibitem[Gao \latin{et~al.}(2005)Gao, Zou, Liu, Xiao, Peng, Song, and
  Liu]{Liu05_054102}
Gao,~J.; Zou,~W.; Liu,~W.; Xiao,~Y.; Peng,~D.; Song,~B.; Liu,~C. Time-Dependent
  Four-Component Relativistic Density-Functional Theory for Excitation
  Energies. II. The Exchange-Correlation Kernel. \emph{J. Chem. Phys.}
  \textbf{2005}, \emph{123}, 054102\relax
\mciteBstWouldAddEndPuncttrue
\mciteSetBstMidEndSepPunct{\mcitedefaultmidpunct}
{\mcitedefaultendpunct}{\mcitedefaultseppunct}\relax
\EndOfBibitem
\bibitem[Wang \latin{et~al.}(2005)Wang, Ziegler, van Lenthe, van Gisbergen, and
  Baerends]{Baerends05_204103}
Wang,~F.; Ziegler,~T.; van Lenthe,~E.; van Gisbergen,~S.; Baerends,~E.~J. The
  calculation of excitation energies based on the relativistic two-component
  zeroth-order regular approximation and time-dependent density-functional with
  full use of symmetry. \emph{J. Chem. Phys.} \textbf{2005}, \emph{122},
  204103\relax
\mciteBstWouldAddEndPuncttrue
\mciteSetBstMidEndSepPunct{\mcitedefaultmidpunct}
{\mcitedefaultendpunct}{\mcitedefaultseppunct}\relax
\EndOfBibitem
\bibitem[Salek \latin{et~al.}(2005)Salek, Helgaker, and Saue]{Saue05_187}
Salek,~P.; Helgaker,~T.; Saue,~T. Linear Response at the 4-component
  Relativistic Density-functional Level: Application to the Frequency-dependent
  Dipole Polarizability of {Hg}, {AuH} and {PtH{$_{2}$}}. \emph{Chem. Phys.}
  \textbf{2005}, \emph{311}, 187--201\relax
\mciteBstWouldAddEndPuncttrue
\mciteSetBstMidEndSepPunct{\mcitedefaultmidpunct}
{\mcitedefaultendpunct}{\mcitedefaultseppunct}\relax
\EndOfBibitem
\bibitem[Casarin \latin{et~al.}(2007)Casarin, Finetti, Vittadini, Wang, and
  Ziegler]{Ziegler07_5270}
Casarin,~M.; Finetti,~P.; Vittadini,~A.; Wang,~F.; Ziegler,~T. {Spin-Orbit
  Relativistic Time-Dependent Density Functional Calculations of the Metal and
  Ligand Pre-Edge XAS Intensities of Organotitanium Complexes: TiCl$_4$,
  Ti($\eta^5$-C$_5$H$_5$)Cl$_3$, and Ti($\eta^5$-C$_5$H$_5$)$_2$Cl$_2$}.
  \emph{J. Phys. Chem. A} \textbf{2007}, \emph{111}, 5270--5279\relax
\mciteBstWouldAddEndPuncttrue
\mciteSetBstMidEndSepPunct{\mcitedefaultmidpunct}
{\mcitedefaultendpunct}{\mcitedefaultseppunct}\relax
\EndOfBibitem
\bibitem[Peralta \latin{et~al.}(2007)Peralta, Scuseria, and
  Frisch]{Frisch07_125119}
Peralta,~J.~E.; Scuseria,~G.~E.; Frisch,~M.~J. Noncollinear Magnetism in
  Density Functional Calculations. \emph{Phys. Rev. B} \textbf{2007},
  \emph{75}, 125119\relax
\mciteBstWouldAddEndPuncttrue
\mciteSetBstMidEndSepPunct{\mcitedefaultmidpunct}
{\mcitedefaultendpunct}{\mcitedefaultseppunct}\relax
\EndOfBibitem
\bibitem[Bast \latin{et~al.}(2009)Bast, Jensen, and Saue]{Saue09_2091}
Bast,~R.; Jensen,~H. J.~A.; Saue,~T. Relativistic Adiabatic Time-Dependent
  Density Functional Theory Using Hybrid Functionals and Noncollinear Spin
  Magnetization. \emph{Int. J. Quant. Chem.} \textbf{2009}, \emph{109},
  2091--2112\relax
\mciteBstWouldAddEndPuncttrue
\mciteSetBstMidEndSepPunct{\mcitedefaultmidpunct}
{\mcitedefaultendpunct}{\mcitedefaultseppunct}\relax
\EndOfBibitem
\bibitem[Devarajan \latin{et~al.}(2009)Devarajan, Gaenko, and
  Autschbach]{Autschbach09_194102}
Devarajan,~A.; Gaenko,~A.; Autschbach,~J. Two-Component Relativistic Density
  Functional Method for Computing Nonsingular Complex Linear Response of
  Molecules Based on the Zeroth Order Regular Approximation. \emph{J. Chem.
  Phys.} \textbf{2009}, \emph{130}, 194102\relax
\mciteBstWouldAddEndPuncttrue
\mciteSetBstMidEndSepPunct{\mcitedefaultmidpunct}
{\mcitedefaultendpunct}{\mcitedefaultseppunct}\relax
\EndOfBibitem
\bibitem[Scalmani and Frisch(2012)Scalmani, and Frisch]{Frisch12_2193}
Scalmani,~G.; Frisch,~M.~J. A New Approach to Noncollinear Spin Density
  Functional Theory Beyond the Local Density Approximation. \emph{J. Chem.
  Theory Comput.} \textbf{2012}, \emph{8}, 2193\relax
\mciteBstWouldAddEndPuncttrue
\mciteSetBstMidEndSepPunct{\mcitedefaultmidpunct}
{\mcitedefaultendpunct}{\mcitedefaultseppunct}\relax
\EndOfBibitem
\bibitem[Bulik \latin{et~al.}(2013)Bulik, Scalmani, Frisch, and
  Scuseria]{Scuseria13_035117}
Bulik,~I.~W.; Scalmani,~G.; Frisch,~M.~J.; Scuseria,~G.~E. Noncollinear Density
  Functional Theory Having Proper Invariance and Local Torque Properties.
  \emph{Phys. Rev. B} \textbf{2013}, \emph{87}, 035117\relax
\mciteBstWouldAddEndPuncttrue
\mciteSetBstMidEndSepPunct{\mcitedefaultmidpunct}
{\mcitedefaultendpunct}{\mcitedefaultseppunct}\relax
\EndOfBibitem
\bibitem[K\"{u}hn and Weigend(2013)K\"{u}hn, and Weigend]{Weigend13_5341}
K\"{u}hn,~M.; Weigend,~F. Implementation of Two-Component Time-Dependent
  Density Functional Theory in TURBOMOLE. \emph{J. Chem. Theory Comput.}
  \textbf{2013}, \emph{9}, 5341--5348\relax
\mciteBstWouldAddEndPuncttrue
\mciteSetBstMidEndSepPunct{\mcitedefaultmidpunct}
{\mcitedefaultendpunct}{\mcitedefaultseppunct}\relax
\EndOfBibitem
\bibitem[Li \latin{et~al.}(2013)Li, Suo, Zhang, Xiao, and Liu]{Liu13_3741}
Li,~Z.; Suo,~B.; Zhang,~Y.; Xiao,~Y.; Liu,~W. Combining Spin-adapted Open-shell
  TD-DFT with Spin-orbit Coupling. \emph{Mol. Phys.} \textbf{2013}, \emph{111},
  3741--3755\relax
\mciteBstWouldAddEndPuncttrue
\mciteSetBstMidEndSepPunct{\mcitedefaultmidpunct}
{\mcitedefaultendpunct}{\mcitedefaultseppunct}\relax
\EndOfBibitem
\bibitem[Eich and Gross(2013)Eich, and Gross]{Gross13_156401}
Eich,~F.; Gross,~E. Transverse Spin-gradient Functional for Noncollinear
  Spin-density-functional Theory. \emph{Phys. Rev. Lett.} \textbf{2013},
  \emph{111}, 156401\relax
\mciteBstWouldAddEndPuncttrue
\mciteSetBstMidEndSepPunct{\mcitedefaultmidpunct}
{\mcitedefaultendpunct}{\mcitedefaultseppunct}\relax
\EndOfBibitem
\bibitem[Eich \latin{et~al.}(2013)Eich, Pittalis, and
  Vignale]{Vignale13_245102}
Eich,~F.; Pittalis,~S.; Vignale,~G. Transverse and Longitudinal Gradients of
  the Spin Magnetization in Spin-density-functional Theory. \emph{Phys. Rev. B}
  \textbf{2013}, \emph{88}, 245102\relax
\mciteBstWouldAddEndPuncttrue
\mciteSetBstMidEndSepPunct{\mcitedefaultmidpunct}
{\mcitedefaultendpunct}{\mcitedefaultseppunct}\relax
\EndOfBibitem
\bibitem[Goings \latin{et~al.}(2018)Goings, Egidi, and Li]{Li18_e25398}
Goings,~J.~J.; Egidi,~F.; Li,~X. Current Development of Non-collinear
  Electronic Structure Theory. \emph{Int. J. Quant. Chem.} \textbf{2018},
  \emph{118}, e25398\relax
\mciteBstWouldAddEndPuncttrue
\mciteSetBstMidEndSepPunct{\mcitedefaultmidpunct}
{\mcitedefaultendpunct}{\mcitedefaultseppunct}\relax
\EndOfBibitem
\bibitem[Dyall and {F{\ae}gri, Jr.}(2007)Dyall, and {F{\ae}gri,
  Jr.}]{Faegri07_book}
Dyall,~K.~G.; {F{\ae}gri, Jr.},~K. \emph{Introduction to Relativistic Quantum
  Chemistry}; Oxford University Press, 2007\relax
\mciteBstWouldAddEndPuncttrue
\mciteSetBstMidEndSepPunct{\mcitedefaultmidpunct}
{\mcitedefaultendpunct}{\mcitedefaultseppunct}\relax
\EndOfBibitem
\bibitem[Jefimenko(1989)]{Jefimenko_book}
Jefimenko,~O.~D. \emph{Electricity and magnetism: An introduction to the theory
  of electric and magnetic fields}; Electret Scientific, 1989\relax
\mciteBstWouldAddEndPuncttrue
\mciteSetBstMidEndSepPunct{\mcitedefaultmidpunct}
{\mcitedefaultendpunct}{\mcitedefaultseppunct}\relax
\EndOfBibitem
\bibitem[Markus(1979)]{Markus_book}
Markus,~Z. \emph{Electromagnetic Field Theory: a problem solving approach}; R.
  F. Krieger Reprint ed.: Malabar, 1979\relax
\mciteBstWouldAddEndPuncttrue
\mciteSetBstMidEndSepPunct{\mcitedefaultmidpunct}
{\mcitedefaultendpunct}{\mcitedefaultseppunct}\relax
\EndOfBibitem
\bibitem[Georg and Furthmüller(1996)Georg, and
  Furthmüller]{Furthmuller96_11169}
Georg,~K.; Furthmüller,~J. Efficient Iterative Schemes for Ab Initio
  Total-energy Calculations Using a Plane-wave Basis Set. \emph{Phys. Rev. B}
  \textbf{1996}, \emph{54}, 11169\relax
\mciteBstWouldAddEndPuncttrue
\mciteSetBstMidEndSepPunct{\mcitedefaultmidpunct}
{\mcitedefaultendpunct}{\mcitedefaultseppunct}\relax
\EndOfBibitem
\bibitem[Georg and Joubert(1999)Georg, and Joubert]{Joubert99_1758}
Georg,~K.; Joubert,~D. From Ultrasoft Pseudopotentials to the Projector
  Augmented-wave Method. \emph{Phys. Rev. B} \textbf{1999}, \emph{59},
  1758\relax
\mciteBstWouldAddEndPuncttrue
\mciteSetBstMidEndSepPunct{\mcitedefaultmidpunct}
{\mcitedefaultendpunct}{\mcitedefaultseppunct}\relax
\EndOfBibitem
\bibitem[Perdew \latin{et~al.}(1996)Perdew, P., Burke, and
  Ernzerhof]{Ernzerhof96_3865}
Perdew; P.,~J.; Burke,~K.; Ernzerhof,~M. Generalized Gradient Approximation
  Made Simple. \emph{Phys. Rev. Lett.} \textbf{1996}, \emph{77}, 3865\relax
\mciteBstWouldAddEndPuncttrue
\mciteSetBstMidEndSepPunct{\mcitedefaultmidpunct}
{\mcitedefaultendpunct}{\mcitedefaultseppunct}\relax
\EndOfBibitem
\bibitem[Stefan \latin{et~al.}(2010)Stefan, Antony, Ehrlich, and
  Krieg]{Krieg10_15}
Stefan,~G.; Antony,~J.; Ehrlich,~S.; Krieg,~H. A Consistent and Accurate Ab
  Initio Parametrization of Density Functional Dispersion Correction (DFT-D)
  for the 94 Elements H-Pu. \emph{J. Chem. Phys.} \textbf{2010}, \emph{132},
  15\relax
\mciteBstWouldAddEndPuncttrue
\mciteSetBstMidEndSepPunct{\mcitedefaultmidpunct}
{\mcitedefaultendpunct}{\mcitedefaultseppunct}\relax
\EndOfBibitem
\bibitem[Stefan \latin{et~al.}(2011)Stefan, Ehrlich, and
  Goerigk]{Goerigk11_1456}
Stefan,~G.; Ehrlich,~S.; Goerigk,~L. Effect of the Damping Function in
  Dispersion Corrected Density Functional Theory. \emph{J. Chem. Phys.}
  \textbf{2011}, \emph{32}, 1456--1465\relax
\mciteBstWouldAddEndPuncttrue
\mciteSetBstMidEndSepPunct{\mcitedefaultmidpunct}
{\mcitedefaultendpunct}{\mcitedefaultseppunct}\relax
\EndOfBibitem
\bibitem[Williams-Young \latin{et~al.}(2020)Williams-Young, Petrone, Sun,
  Stetina, Lestrange, Hoyer, Nascimento, Koulias, Wildman, Kasper, Goings,
  Ding, DePrince~III, Valeev, and Li]{Li20_e1436}
Williams-Young,~D.~B.; Petrone,~A.; Sun,~S.; Stetina,~T.~F.; Lestrange,~P.;
  Hoyer,~C.~E.; Nascimento,~D.~R.; Koulias,~L.; Wildman,~A.; Kasper,~J.;
  Goings,~J.~J.; Ding,~F.; DePrince~III,~A.~E.; Valeev,~E.~F.; Li,~X. The
  Chronus Quantum (ChronusQ) Software Package. \emph{WIREs Comput. Mol. Sci.}
  \textbf{2020}, \emph{10}, e1436\relax
\mciteBstWouldAddEndPuncttrue
\mciteSetBstMidEndSepPunct{\mcitedefaultmidpunct}
{\mcitedefaultendpunct}{\mcitedefaultseppunct}\relax
\EndOfBibitem
\bibitem[Li \latin{et~al.}(2018)Li, Williams-Young, Valeev, Petrone, Sun,
  Stetina, and Kasper]{chronusq_beta2}
Li,~X.; Williams-Young,~D.; Valeev,~E.~F.; Petrone,~A.; Sun,~S.; Stetina,~T.;
  Kasper,~J. Chronus Quantum, Beta 2 Version. 2018;
  {http://www.chronusquantum.org}\relax
\mciteBstWouldAddEndPuncttrue
\mciteSetBstMidEndSepPunct{\mcitedefaultmidpunct}
{\mcitedefaultendpunct}{\mcitedefaultseppunct}\relax
\EndOfBibitem
\end{mcitethebibliography}

\end{document}